\documentclass[reqno, a4paper]{amsart}

\usepackage{fullpage, color, hyperref}
\usepackage{latexsym}
\usepackage{amsfonts}
\usepackage{amsmath}
\usepackage{amssymb}
\usepackage{amsthm}
\usepackage{graphicx}
\usepackage{tikz}
\usepackage{array}
\usepackage{caption}
\usetikzlibrary{calc,fit,matrix,arrows,automata,positioning}

\newcounter{example}

\title{Stability of Periodic Travelling Wave Solutions to the Kawahara Equation}

\author[O. Trichtchenko]{Olga Trichtchenko}
\address[O. Trichtchenko]{Department of Physics and Astronomy, University of Western Ontario, London, ON,  N6A 3K7, Canada}
\email{olga.trichtchenko@gmail.com}
\author[B. Deconinck]{Bernard Deconinck}
\address[B. Deconinck]{Department of Applied Mathematics, University of Washington, Seattle, WA, 98195, USA}
\author[R. Koll{\'a}r]{Richard Koll{\'a}r}
\address[R. Koll{\'a}r]{Department of Applied Mathematics and Statistics, Comenius University, 842 48 Bratislava, Slovakia}

\begin{document}

\maketitle

\begin{abstract}
We analyse the stability of periodic, travelling-wave solutions to the Kawahara equation and some of its generalizations. We determine the parameter regime for which these solutions can exhibit resonance. By examining perturbations of small-amplitude solutions, we show that generalised resonance is a mechanism for high-frequency instabilities. We derive a quadratic equation which fully determines the stability region for these solutions. Focussing on perturbations of the small-amplitude solutions, we obtain asymptotic results for how their instabilities develop and grow. Numerical computation is used to confirm these asymptotic results and illustrate regimes where our asymptotic analysis does not apply.
\end{abstract}

\section{Introduction}

The goal of this work is to examine the stability of periodic travelling-wave solutions to the lowest-order dispersive nonlinear scalar partial differential equations (PDEs) that may exhibit instabilities. To this end, we consider a general fifth-order Korteweg-de Vries type-equation of the form
\begin{align}
u_t = \alpha u_{xxx} + \beta u_{5x} + \sigma \left( u^{p+1} \right)_x,
\label{eq:genForm}
\end{align}
with the linear dispersion relation $\omega(k) = \alpha k^3 - \beta k^5$. Here $\alpha, \beta, \sigma$ are parameters with the exponent $p$ taking on positive integer values. We focus mainly on $p=1$ which is the Kawahara equation \cite{K72} (sometimes referred to as the Super KdV equation \cite{HB88}), and $p=2$ which is a modified fifth-order KdV \cite{MGK68}. The system \eqref{eq:genForm}  is Hamiltonian,
\begin{align}
u_t = \partial_x\frac{\delta H}{\delta u},
\label{eq:eqHamil}
\end{align}
with
\begin{align}
H = \int_0^{L} \left( \frac{1}{2} \left( -\alpha u_x^2 + \beta u_{xx}^2 \right) + \frac{\sigma}{p+2} u^{p+2} \right) dx,
\end{align}
and we have restricted ourselves to the finite domain $x\in[0,L]$. Thus, \eqref{eq:eqHamil} provides a convenient setting to use the spectral stability theory of \cite{DT15}. In this manuscript, we derive criteria for instability of small-amplitude periodic solutions of \eqref{eq:eqHamil} and we show how the numerical results match perturbative approximations stemming from the theory. The focus of this work is on high-frequency instabilities, away from the origin in the spectral plane. In other words, we do not discuss the modulational or Benjamin-Feir instability.

The stability theory for the third-order KdV and mKdV equations and their generalizations with different nonlinearities is well established, both for solutions defined on the whole line and for periodic solutions, see for instance \cite{kapituladeconinck} and references therein.  In contrast, the literature for stability studies of the Kawahara equation and its generalizations is limited.
For whole-line solutions, Bridges {\em et al.} \cite{BDG02} devise a computational method using Evans functions to examine the stability of solitary-wave solutions to fifth-order KdV with polynomial nonlinearities and they show where instabilities arise. Haragus {\em et al.} \cite{HLS06} consider stability of periodic solutions to the Kawahara equation. They conclude that waves whose amplitude scales as the $5/4$-th power of the wave speed are stable. Our results are not restricted to this scaling regime, nor are we limited to the specific Kawahara nonlinearity.
Our focus is on the spectral stability of periodic small-amplitude travelling waves of \eqref{eq:eqHamil}. In particular, we examine how these waves behave if they are perturbed by high-frequency disturbances of any period. We start by applying the theory of \cite{DT15}, where we have shown that third-order KdV does not exhibit high-frequency instabilities.

Our problem falls within the general class of problems studied by Haragus \& Kapitula \cite{HK08} and Johnson {\em et al.} \cite{JZB10}, or computationally, Deconinck \& Kutz \cite{deconinck1}: a Floquet-Bloch decomposition is used to decompose the instability spectrum (consisting of a collection of curves) into a union of point spectra, corresponding to perturbations with a specific Floquet exponent, see below.

The equations we analyse have two dispersive terms, which depending on the sign of $\alpha$ and $\beta$, allows for two linear waves with different wavenumbers to travel at the same speed. Since such equations are often used to describe water waves in the long-wave regime where the forces of gravity and surface tension are both important \cite{K72, SW02}, our study gives insight into the mechanism for instability in the context of more complicated equations describing water waves. It is known that equations admitting bidirectional waves can exhibit high-frequency instabilities \cite{DT15, CJ17} and in this work, we show that resonance provides another mechanism, even in the context of one-directional wave propagation.

Resonant phenomena are not only interesting from a stability perspective, but they also affect the asymptotic analysis of solutions. Haupt \& Boyd \cite{HB88} showed how to modify the series representation (Stokes expansion) of a resonant or near-resonant solution. They presented numerical results near the resonant regime, discussing how resonance affects the ordering of the coefficients for the asymptotic series expansion. Akers \& Gao \cite{AG12} considered near-bichromatic solutions to models with a quadratic nonlinearity and a general dispersion relation. We consider a dispersion relation containing both third- and fifth-order terms, while allowing for a general nonlinearity. We construct an asymptotic form of the solutions near and away from resonance and we show how the asymptotic representation of the solutions indicates stability properties depending on the regime the solution is in.

There is a vast amount of literature on the stability of solutions in the context of the full Euler equations and only the most relevant references are discussed here. Eigenvalues of the spectral stability problem for the water wave problem using an approach that takes advantage of its Hamiltonian nature, are considered by MacKay \& Saffman \cite{MS86}. Away from the resonant regime, it has been noted by McLean \cite{M82} that an instability that can be described by an $N$th order interaction that grows at order $N$. Akers \cite{A15} further discussed the relationship between the $N$th order interaction and the collision of unstable eigenvalues in the spectral stability problem. He devised a perturbation expansion for these eigenvalues and discussed their radius of analyticity illustrating the mechanism for instabilities. For resonant solutions describing water waves, it has been shown that an asymptotic expansion for the solutions follows a similar pattern to the work of Haupt \& Boyd in \cite{TDW16} and that resonance has an effect on the stability of these solutions.

%

The layout of the paper is as follows. In Section \ref{sec:instabCriteria} we derive a necessary condition for instability and discuss how it is impacted by resonance. In Section \ref{sec:asymptotics}, we derive asymptotic approximations to solutions and matrices describing spectral stability, in particular focussing on Kawahara and more general equations. We illustrate numerically how well these asymptotic results work in Section \ref{sec:numerics}. We conclude in Section \ref{sec:conclusion}.

\section{Instability Criteria for Small-Amplitude Solutions}\label{sec:instabCriteria}

We begin by deriving a criterion for the instability of a travelling wave solution, involving $\alpha, \beta$ and $\sigma$, without requiring the functional form of this solution. Using this criterion, we are able to look at a particular form of the solutions and their perturbations in different parameter regimes.
Moving to a frame of reference travelling at speed $V$, we obtain
\begin{align}
u_t = Vu_x + \alpha u_{xxx} + \beta u_{5x} + \sigma \left( u^{p+1} \right)_x,
\label{eq:travGenForm}
\end{align}
Let $u^{(0)}(x)$ be a stationary solution of period $L$ of \eqref{eq:travGenForm} (corresponding to a travelling wave solution of \eqref{eq:genForm}) and $u^{(1)}(x)$ a small perturbation of this solution such that
\begin{align}
u(x,t) = u^{(0)}(x) + \delta e^{\lambda t} u^{(1)}(x) + O(\delta^2),
\label{eq:pertGen}
\end{align}
with $\delta$ small, using separation of variables to justify the time dependence of the first-order term.

At zeroth order in $\delta$,
the steady-state problem for a travelling wave with speed $V$ is given by
\begin{align}
Vu^{(0)}_x + \alpha u^{(0)}_{xxx} + \beta u^{(0)}_{5x} + \sigma \left[ (u^{(0)})^{p+1} \right]_x = 0,
\label{eq:travSteadyState}
\end{align}
where $u^{(0)}(x)$ is periodic with period $L$. Using the scaling symmetry of the equation, we may choose $L=2\pi$, so that
\begin{align}
u^{(0)}(x) = \sum_{n=-\infty}^{\infty}\hat{u}^{(0)}_k e^{ikx}.
\label{eq:u0Form}
\end{align}

At first order in $\delta$, using \eqref{eq:pertGen} in \eqref{eq:travGenForm} gives 
\begin{align}
\lambda u^{(1)}  = Vu^{(1)}_x + \alpha u^{(1)}_{xxx} & + \beta u^{(1)}_{5x} + \sigma (p+1)  \left[ \left(u^{(0)}\right)^{p} u^{(1)} \right]_x.
\label{eq:firstOrderGen}
\end{align}
We do not restrict the perturbation $u^{(1)}(x)$ to have the same period as $u^{(0)}(x)$. By Floquet's theorem \cite{deconinck1}, all bounded solutions of \eqref{eq:firstOrderGen} are of the form
\begin{align}
u^{(1)}(x) = e^{i\mu x}\sum_{m=-\infty}^{\infty} \hat{u}^{(1)}_m e^{imx} + c.c.,
\label{eq:u1Form}
\end{align}
where $\mu \in [0,1/2)$ is the Floquet parameter and $c.c.$ denotes the complex conjugate. Equation \eqref{eq:firstOrderGen} is a spectral problem where  eigenfunctions corresponding to $\lambda$ with Re($\lambda$)$>0$ give rise to unstable perturbations $u^{(1)}(x)$.

For small-amplitude waves, the nonlinear term in \eqref{eq:firstOrderGen} can be neglected. This implies that our perturbative analysis applies regardless of the exponent of the nonlinearity. Using \eqref{eq:u0Form} and omitting the nonlinear term, the coefficient of $\hat{u}^{(0)}_k$ in the steady-state problem  \eqref{eq:travSteadyState} is
\begin{align}
ikV_0 - ik^3 \alpha + ik^5\beta = 0,
\label{eq:smallAmpBifurc}
\end{align}
so that
\begin{align}
V_0 = k^2 \alpha - k^4 \beta.
\end{align}
Here $V_0$ is the leading-order term in the $\delta$-expansion of $V$. We choose the first Fourier coefficient $\hat{u}^{(0)}_1$ as a free parameter, so that $k=1$ and
$V_0 = \alpha - \beta$. This gives the bifurcation point $(0,V_0)$ in the $(\hat{u}^{(0)}_1,V)$-plane from which non-zero solutions
emanate. However, if for $k=K\neq 1$
\begin{align}
\beta = \frac{\alpha}{K^2+1},
\label{eq:resBeta}
\end{align}
then the two modes with wavenumbers $k=1$ and $k=K$ travel with the same speed and there are two free coefficients $\hat{u}^{(0)}_1$ and $\hat{u}^{(0)}_K$ in \eqref{eq:u0Form}. This is referred to as resonance.

Using \eqref{eq:u1Form}, the spectral problem \eqref{eq:firstOrderGen} in Fourier space to leading order in $\delta$ is
\begin{align}
\lambda_m^{\mu} = i(m+\mu)V_0 - i (m+\mu)^3\alpha & + i (m+\mu)^5\beta,
\label{eq:smallAmpGenEval}
\end{align}
leading to a purely imaginary spectrum and the conclusion that the zero solution is spectrally stable since perturbations in \eqref{eq:pertGen} do not grow exponentially in time. Spectrally unstable perturbations require Re$(\lambda)>0$. Since \eqref{eq:genForm} is Hamiltonian, the spectrum of \eqref{eq:travGenForm} is symmetric with respect to both the real and imaginary $\lambda$ axis, and for every element of the spectrum with Re$(\lambda)>0$, there is another one for which Re$(\lambda)<0$. The spectrum depends continuously on the amplitude of the solution. As the amplitude increases, a pair of purely imaginary eigenvalues may collide, after which they can leave the imaginary axis symmetrically. This requires that for a given perturbation with Floquet parameter $\mu$, there is a pair $(m,n)$ such that
\begin{align}
\lambda^{\mu}_m = \lambda^{\mu}_n\in i\mathbb{R},
\label{eq:genCollision}
\end{align}
The location of these eigenvalues varies with the amplitude of the solution as we move up the solution bifurcation branch.

For fixed $\alpha$, we can ensure a $\lambda_n^\mu$ and $\lambda_m^\mu$ collide by choosing
\begin{align}
\beta = \alpha\frac{(m+\mu)(1-(m+\mu)^2)-(n+\mu)(1-(n+\mu)^2)}{(m+\mu)(1-(m+\mu)^4)-
(n+\mu)(1-(n+\mu)^4)},
\label{eq:betaCollision}
\end{align}
with bifurcation branch starting at $V_0 = \alpha-\beta$.
We refer to this as a generalised resonance condition: a resonance between modes with wave numbers $\mu+m$ and $\mu+n$, which are not restricted to be $2\pi$ periodic. The regular resonance condition \eqref{eq:resBeta} is obtained from \eqref{eq:betaCollision} by imposing $\mu=0$ and $m=1$.
We conclude that the only mechanism for an instability to occur for a small-amplitude solution for the fifth-order KdV, is due to the presence of the parameter $\beta$, leading to a generalised resonance.

In general, it is easier to check for eigenvalue collisions without imposing $\mu \in [0, 1/2)$.  Instead we consider the ``unfolded'' version of the collision condition given by
\begin{align}
\lambda_0^{\mu} = \lambda^{\mu}_{|m-n|}\in i \mathbb{R}.
\label{eq:unfoldCollision}
\end{align}
We emphasize that this condition depends on the difference between the Fourier modes of the perturbation, i.e., on $|m-n|$. In order for an instability to occur, the condition
\begin{align}
5 \beta \mu^4 + 10 \beta \bar{n} \mu^3 + (10\beta\bar{n}^2 - 3\alpha )\mu^2 + (5\beta \bar{n}^3 - 3\alpha \bar{n})\mu  + \beta \bar{n}^4  - \alpha\bar{n}^2  + V_0 = 0,
\label{eq:quartic}
\end{align}
must be satisfied. This is obtained from \eqref{eq:unfoldCollision} by substituting the form of $\lambda$ from  \eqref{eq:smallAmpGenEval}, with $\bar{n} = |m-n|$. This simplifies to a fourth-order polynomial in $\mu$. This is a condition for collisions to occur at zero amplitude. Even if this condition is not satisfied for a particular choice of $\alpha$ and $\beta$, instabilities may result from  eigenvalue collisions for non-zero amplitude solutions along the bifurcation branch. In what follows, we replace $\bar n$ by $n$, to ease notation.

\begin{figure}
\begin{center}
\includegraphics[width=0.6\textwidth]{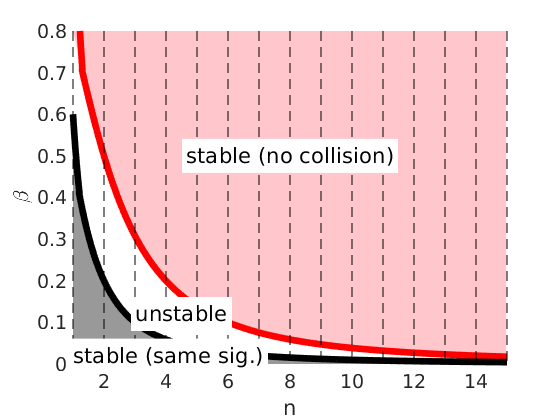}
\caption{The coloured regions represent the areas in the $\beta - n$ space for which there are no unstable collisions of eigenvalues.\label{fig:totInstab}}
\end{center}
\end{figure}

As the amplitude is increased from zero, collided eigenvalues move in a way that conserves the total energy (Hamiltonian) of the system. For eigenvalues to move off the imaginary axis, they need to have opposite sign contributions to the Hamiltonian \cite{DT15}, i.e.,  their Krein signatures are opposite \cite{krein2,MS86}. For solutions of small amplitude of a system of the form \eqref{eq:eqHamil}, it is straightforward to compute the Krein signatures $\kappa(\lambda_n^\mu)$, $\lambda_n^\mu\neq 0$ \cite{DT15}:
\begin{align}
\kappa\left( \lambda_{n}^{\mu}\right) = \text{sgn}\left(\frac{\omega(n+\mu)}{n+\mu}\right).
\end{align}
It follows \cite{DT15} that colliding eigenvalues $\lambda_0^{\mu}$ and
$\lambda^{\mu}_{n}$ have opposite Krein signatures if and only if
\begin{align}
s = \mu(\mu + n) < 0.
\label{eq:c}
\end{align}
In \cite{KDT18}, we prove a general result that shows the quartic equation \eqref{eq:quartic} is rewritten in terms of $s$ as
\begin{align}
5 \beta s^2 + (5\beta n^2 - 3 \alpha)s + \beta n^4 - \alpha n^2 + V_0 = 0.
\label{eq:quadratic}
\end{align}
Without loss of generality, we let $\alpha = 1$ and we focus on the $V_0=\alpha - \beta=1-\beta$ solution branch. The criteria for both roots of the above quadratic equation to be negative is given by \cite{KDT18}
\begin{align}
\beta > \max{\left(\frac{3}{5n^2}, \frac{1}{n^2+1}\right)},
\label{eq:minBound}
\end{align}
in which case the colliding eigenvalues have opposite signature. Furthermore, the bound above which there are no collisions is \cite{DT15}
\begin{align}
\beta < \min\left( \frac{6}{5n^2}, \frac{1}{(n/2)^2+1} \right).
\label{eq:maxBound}
\end{align}
The stability region given by the bounds \eqref{eq:minBound} and \eqref{eq:maxBound} is shown in Figure~\ref{fig:totInstab}.

\section{Asymptotic Analysis}\label{sec:asymptotics}

We examine the possible instability regions for a non-zero amplitude solution. Small-amplitude solutions to generalised KdV equations are straightforward to compute perturbatively. First we integrate the zeroth order, steady-state equation \eqref{eq:travSteadyState} to obtain
\begin{align}
Vu^{(0)}  + \alpha u^{(0)}_{xx} + \beta u^{(0)}_{4x} + \sigma (u^{(0)})^{p+1} + \text{const.} = 0.
\label{eq:zeroInt}
\end{align}
Using \eqref{eq:u0Form}, reality of the solutions requires $\hat{u}^{(0)}_{k} = u^{(0)}_{-k}$.  The solution is constructed as a cosine series:
\begin{align}
u^{(0)} = a_0 + \sum_{n=1}^{\infty} a_n \cos(nx).
\label{eq:cosSeries}
\end{align}
As mentioned above, we use $u^{(0)}_{1} = a_1$ as a small parameter so that the solution contains the $k=1$ mode. The integration constant in \eqref{eq:zeroInt} may be equated to zero or, alternatively, we may choose $a_{0} = 0$.

With the solution in hand, we substitute \eqref{eq:u1Form} into \eqref{eq:firstOrderGen} to examine the resulting stability spectral problem
\begin{align}
\lambda \sum_{m=-\infty}^{\infty} \hat{u}^{(1)}_m e^{imx} = i \sum_{m=-\infty}^{\infty}\left[ V(m+\mu)- \alpha(m+\mu)^3 + \beta(m+\mu)^5\right]  \hat{u}^{(1)}_m e^{imx} \nonumber \\
+ \sigma (p+1) \partial_x\left[(u^{(0)}(x))^{p} \sum_{m=-\infty}^{\infty} \hat{u}^{(1)}_m e^{imx}\right],
\label{eq:firstOrderSeries}
\end{align}
Multiplying by $e^{-ikx}$ and integrating over one period,
\begin{align}
\lambda\hat{u}^{(1)}_k  = i \left[V (k+\mu)- \alpha(k+\mu)^3 + \beta(k+\mu)^5\right]  \hat{u}^{(1)}_k
+ \sigma \frac{p\!+\!1}{2\pi} \!\int_{0}^{2\pi} \partial_x\left[(u^{(0)}(x))^{p}\!\!\!\sum_{m=-M}^{M} \hat{u}^{(1)}_m e^{imx}\right]e^{-ikx} dx
\label{eq:firstOrderInt}
\end{align}
We focus on two different cases for the nonlinearity, $p=1$ and $p=2$. This allows us to write down the form of the matrix entries for the stability problem explicitly in terms of the coefficients of \eqref{eq:cosSeries}. From Figure \ref{fig:totInstab}, we see that for sufficiently small $\beta$, there is a limited number ${n}$ of modes giving rise to eigenvalue collisions which can lead to instabilities through Hamilton-Hopf bifurcations. In order to capture all possible instabilities, it is reasonable to truncate the Fourier series expansion \eqref{eq:u1Form} at $M$ modes, with $2M \ge {n}$. Since only pairwise eigenvalue collisions are considered, two modes contribute to each collision. We can isolate these modes and examine the resulting $2\times 2$ matrix. To get more accurate results, we can consider increasingly larger matrices since \eqref{eq:cosSeries} is valid for small range of amplitudes only. The region of validity shrinks as ${n}$ increases.

\subsection{Kawahara Equation: $p=1$}
The series solution \eqref{eq:cosSeries} for \eqref{eq:travSteadyState} with $p=1$ is obtained from the recurrence relation
\begin{align}
(V - \alpha k^2 + \beta k^4)a_k = -\frac{\sigma}{2}\sum_{n=k}^{\infty}a_na_{n-k}- \frac{\sigma}{2}\sum_{n=0}^ka_na_{k-n}.
\label{eq:kawaharaKeq}
\end{align}
As stated above, we construct a formal perturbation series for the solution. We restrict to small-amplitude waves. Since the solution bifurcates away from the zero amplitude solution at $V = V_0 = \alpha - \beta$, we let $a_1 = \epsilon$, with $\epsilon$ small. By dominant balance we obtain at leading order
\begin{align}
a_0 & = -\frac{\sigma}{2}\frac{1}{V_0}a_1^2 + O(\epsilon^3), \\
a_2 & = -\frac{\sigma}{2}\frac{1}{V_0-2^2\alpha+2^4\beta}a_1^2  + O(\epsilon^3),\\
a_3 & = -\frac{\sigma}{2}\frac{1}{V_0-3^2\alpha+3^4\beta}(2a_2a_1)  + O(\epsilon^4), \\
a_4 & = -\frac{\sigma}{2}\frac{1}{V_0-4^2\alpha+4^4\beta}(a_2^2 + 2a_3a_1) +  O(\epsilon^5),
\end{align}
and so on, where we need to ensure the resonance condition $V_0-k^2\alpha+k^4\beta = 0$ is not satisfied for $k=2,3,4$ so the perturbation series is well ordered.

Note that for the above to work at all orders, it is necessary that $V$ is expanded as a series in $\epsilon$ as well:
\begin{align}
V = \sum_{n=0}^{\infty} \epsilon^n V_n,
\label{eq:waveForm}
\end{align}
with $V_0 = \alpha-\beta$. For our purposes, the explicit form of the other terms is not needed. From the equations for the Fourier coefficients $a_n$, it is easy to see that modes $a_n$ with odd index $n$ contain only odd powers of $\epsilon$, because we are grouping coefficients multiplying $e^{ikx}$ using a quadratic nonlinearity. It follows from \eqref{eq:kawaharaKeq}  that ${V}$ can only contain even powers of $\epsilon$.

We normalize the solution so that $a_0 = 0$ by adjusting $V$ using \eqref{eq:cosSeries} in \eqref{eq:zeroInt}. This allows us to obtain the expression for $a_k$ by rearranging \eqref{eq:kawaharaKeq} as
\begin{align}
a_k = -\frac{\sigma}{2}\frac{1}{V_0-k^2\alpha+k^4\beta}c_k \epsilon^k + O(\epsilon^{k+1}),
\end{align}
where $c_k$ is independent of $\epsilon$. We note that in the case of resonance for $k=K$, the coefficient $a_K$ is of lower order than $\epsilon^K$ (\textit{i.e.}, this coefficient is more important than in the non-resonant case). We show this numerically in Section~4.

For the Kawahara equation ($p=1$), the stability spectral problem \eqref{eq:firstOrderInt} simplifies to
\begin{align}
\lambda \hat{u}^{(1)}_k  = i \left[V (k+\mu)- \alpha(k+\mu)^3 + \beta(k+\mu)^5\right]  \hat{u}^{(1)}_k + 2\sigma i(\mu+k) \sum_{m=-\infty}^{\infty}  \hat{u}^{(0)}_m \hat{u}^{(1)}_{(k-m)}. \label{eq:nonLinForm}
\end{align}
Denoting the vector of the Fourier coefficients by $\vec{U}^{(1)} = (\hat{u}^{(1)}_{-M}, \cdots, \hat{u}^{(1)}_{-1},\hat{u}^{(1)}_{1}, \cdots, \hat{u}^{(1)}_{M} )^T$, the spectral problem is written as a system in the form
\begin{align}
\lambda \vec{U}^{(1)} = S \vec{U}^{(1)}.
\end{align}
We write the matrix $S$ as
\begin{align}
S = iD+iT,
\label{eq:Smatrix}
\end{align}
where the diagonal matrix $D$ is read off from \eqref{eq:smallAmpGenEval} so that
\begin{align}
d_{m,n} = \begin{cases} (n+\mu)V - (n+\mu)^3\alpha + (n+\mu)^5\beta, \ \
& \text{for} \ \ m=n,\\
0, \ \
& \text{for} \  \ m\neq n.
\end{cases}
\label{eq:Dmatrix}
\end{align}
It follows that the (not necessarily off-diagonal) matrix $T$ is determined by the nonlinearity. It is given by
\begin{align}
t_{m,n} = 2\sigma \begin{cases} 0, \ \ & \text{for} \ \ m=n,\\
(\mu - M + m - 1)a_{|n-m|}, \ \ &\text{for} \  \ m\neq n.
\end{cases}
\label{eq:Tmatrix}
\end{align}
Here $M$ is the number of modes in the truncated expression for the perturbation given by \eqref{eq:u1Form}.
We observe that for a zero-amplitude solution, $T=0$ and $S$ is diagonal as expected. To retain the four-fold symmetry for the eigenvalues of a Hamiltonian system, it is necessary to include the complex conjugate of $u^{(1)}(x)$.

We consider pairwise collisions of eigenvalues. For example, if we choose the parameters $\alpha = 1$ and $\beta = 1/4$, the only real solutions to \eqref{eq:quartic} occur for $\bar{n} = |m-n|\le 3$. In other words, for those parameters we only need $M=2$ to capture all the instabilities since the perturbation has modes $m = -2,-1,1,2$ with the largest difference between modes $|m-n|=4$. More specifically, in order to know the information about collisions between only two modes, we can focus on a $2 \times 2$ matrix $T$. It is worth noting that a collision between modes $m=-2$ and $n=-1$ is equivalent to a collision between modes $m=2$ and $n=1$ due to symmetry. Similarly, a collision between modes $m = -2$ and $n=1$ is the same as between $m = 2$ and $n = -1$ and considering both cases is redundant.

Arguably the most interesting case is that of $(m,n) = (-2,-1)$. We study the perturbation of the eigenvalues by examining the block matrix
\begin{align}
S(-2,-1) = i
\begin{pmatrix}
-V(-2\!+\!\mu) \!+\! \alpha(-2\!+\!\mu)^3 \!-\! \beta(-2\!+\!\mu)^5 & \sigma (\mu-2)a_1 \\
 \sigma(\mu-1)a_1 & -V(-1\!+\!\mu) \!+\! \alpha(-1\!+\!\mu)^3 \!-\! \beta(-1\!+\!\mu)^5  \\
\end{pmatrix}.
\end{align}
We can explicitly compute the eigenvalues $\bar{\lambda}_{-2,-1}^{\mu}$ of $S(-2,-1)$, where the bar denotes that these eigenvalues are perturbations of the eigenvalues in \eqref{eq:smallAmpGenEval}. We obtain
\begin{align}
\bar{\lambda}_{-2,-1}^{\mu} = \frac{i}{2}\left[d_{-1,-1}+d_{-2,-2}\pm\sqrt{(d_{-1,-1}-d_{-2,-2})^2+4\sigma^2 a_1^2(\mu-1)(\mu-2)} \right].
\label{eq:beta1_4col1}
\end{align}
Since $d_{-1,-1}\approx d_{-2,-2}$ near the collision, eigenvalues with non-zero real part are obtained if
\begin{align}
(\mu-2)(\mu-1)<0,
\end{align}
which is equivalent to the Krein signature condition in \cite{DT15}, discussed in more detail in \cite{KDT18}. The exponential growth rate of the instability is proportional to $a_1=O(\epsilon)=O(\epsilon^{|m-n|})$ with $m=-2$ and $n=-1$, consistent with \cite{MS86}.

For the collision between modes $m=-1$ and $n=1$, we consider the matrix
\begin{align}
S{(-1,1)} = i
\begin{pmatrix}
 -V(-1\!+\!\mu) \!+\! \alpha(-1\!+\!\mu)^3 \!-\! \beta(-1\!+\!\mu)^5 &  \sigma(\mu-1)a_2 \\
 \sigma(\mu+1)a_2 & -V(1\!+\!\mu) \!+\! \alpha(1\!+\!\mu)^3 \!-\! \beta(1\!+\!\mu)^5
\end{pmatrix},
\end{align}
with eigenvalues
\begin{align}
\bar{\lambda}_{-1,1}^{\mu} = \frac{i}{2}\left[d_{-1,-1}+d_{1,1}\pm\sqrt{(d_{-1,-1}-d_{1,1})^2+4\sigma^2a_2^2(\mu-1)(\mu+1) } \right].
\label{eq:beta1_4col2}
\end{align}
The condition for the eigenvalue to have a nonzero real part is given by
\begin{align}
(\mu+1)(\mu-1)<0,
\end{align}
which is again equivalent to the Krein signature condition. The exponential part of the growth rate for this instability is proportional to $a_2=O(\epsilon^2)=O(\epsilon^{|m-n|})$ with $|m-n|=2$.

The last case we consider is the collision between modes $m=-2$ and $n=1$.
\begin{align}
S{(-2,1)} = i
\begin{pmatrix}
 -V(-2\!+\!\mu) \!+\! \alpha(-2\!+\!\mu)^3 \!-\! \beta(-2\!+\!\mu)^5 & \sigma (\mu-2)a_3 \\
 \sigma(\mu+1)a_3 & -V(1\!+\!\mu) \!+\! \alpha(1\!+\!\mu)^3 \!-\! \beta(1\!+\!\mu)^5
\end{pmatrix},
\end{align}
with eigenvalues
\begin{align}
\bar{\lambda}_{-2,1}^{\mu} = \frac{i}{2}\left[d_{-2,-2}+d_{1,1}\pm\sqrt{(d_{-2,-2}-d_{1,1})^2+4\sigma^2a_3^2(\mu-2)(\mu+1) } \right].
\label{eq:beta1_4col3}
\end{align}
The condition for the eigenvalues to have a nonzero real part close to their collision is
\begin{align}
(\mu-2)(\mu+1)<0,
\end{align}
equivalent again to the Krein signature condition. The exponential growth rate, i.e., the real part of the eigenvalue, is proportional to $a_3=O(\epsilon^3)=O(\epsilon^{|m-n|})$ with $m=-2$ and $n=1$.

An accurate approximation of the instability growth rate requires the inclusion of all the modes $a_j$. The inclusion of more modes results in more accurate estimates, but becomes more cumbersome as block matrices of increasingly larger size must be dealt with. In practice, it is most relevant to include the modes $a_j$ with $j\in [m,n]$. For instance, for the example of colliding eigenvalues with $m=-2$ and $n=1$, we consider the eigenvalues of a $3\times 3$ matrix
\begin{align}
\bar{S}{(-2,-1,1)} = i
\begin{pmatrix}
 d_{-2,-2} & \sigma(\mu-2)a_1 & (\mu-2)a_3 \\
\sigma(\mu-1)a_1 & d_{-1,-1} & \sigma(\mu-1)a_2 \\
(\mu+1)a_3 & \sigma(\mu+1)a_2 & d_{1,1}
\end{pmatrix}.
\end{align}
In Section \ref{sec:numerics}, we demonstrate the growth rates of the instability as a function of $\epsilon$ numerically for particular parameters. The analytical expression for the eigenvalues are too cumbersome to analyze.

\subsection{Modified Fifth-Order KdV ($p=2$)}
We repeat the above process for \eqref{eq:genForm} with $p=2$. In a moving frame, the stationary problem is
\begin{align}
Vu^{(0)}  + \alpha u^{(0)}_{xx} + \beta u^{(0)}_{4x} = - \sigma (u^{(0)})^3.
\label{eq:cubicSteadyState}
\end{align}
As in the previous section, we truncate at $N=4$ and set $V_0=\alpha-\beta$. We find $a_0 = 0$, $ a_2=0$ and $a_4=0$, giving the same order of approximation as for the quadratic nonlinearity. Further
\begin{align}
a_3 \approx \frac{\sigma}{V - 9\alpha + 81\beta}a_1^3,
\end{align}
provided that $\beta \ne \alpha/10$ (the resonance condition).
As before, to balance the higher-order terms in each of the equations, we introduce an expansion for the wave speed given by \eqref{eq:waveForm}. Since for $k$ even, $a_k = 0$, every odd term in $\bar{V}$ is zero, as in the case $p=1$.

We compute the stability matrix by considering \eqref{eq:firstOrderInt} with $p=2$ and substituting the expansion \eqref{eq:u1Form} for $u^{(1)}$. After performing the appropriate truncations, multiplying by $e^{-ikx}$ and integrating with respect to $x$, we obtain $D$ as in \eqref{eq:Dmatrix} whereas the contribution from the nonlinear term is
\begin{equation}\label{eq:highMatrixMskdv}
T = 3\sigma
\left(\begin{array}{cccc}
2(\mu-2)(a_1^2+a_3^2) & 0 & 0 & 2(\mu-2)a_1a_3 \\
 0 & 2(\mu-1)(a_1^2+a_3^2) & (\mu-1)[a_1^2+2a_1a_3] &  0 \\
0 & (\mu+1)[a_1^2+2 a_1a_3] & 2(\mu+1)(a_1^2+a_3^2) &  0 \\
 2(\mu+2)a_1a_3 &  0 &  0 & 2(\mu+2)(a_1^2+a_3^2)
\end{array}\right).
\end{equation}
The stability matrix for $p=2$ is more sparse than for $p=1$  since the coefficients of the even-order terms in the cosine series for $u^{(0)}$ are zero. We note that there is a non-zero contribution to the diagonal terms for the full matrix $S$.

We can repeat the same analysis as for $p=1$ to determine for which colliding eigenvalues, indexed by $m$ and $n$, we get instabilities. In this case, the analysis is simplified by the presence of many zeros in the matrix $T$. For example, the eigenvalues of the matrix $S({-2,-1})$ are given by
\begin{align}
\bar{\lambda}_{-2,-1}^{\mu} =
\begin{cases}
i \left[ -V(-2 + \mu) + \alpha(-2 + \mu)^3 - \beta(-2 + \mu)^5 + 6\sigma (\mu-2)(a_1^2+a_3^2) \right] \\
i \left[ -V(-1 + \mu) + \alpha(-1 + \mu)^3 - \beta(-1 + \mu)^5 + 6\sigma (\mu-1)(a_1^2+a_3^2) \right]
\end{cases},
\end{align}
and are purely imaginary. This implies the $(-2,-1)$ collision does not result in instabilities, up to this order.
When considering the $(-1,1)$ collision, it is no longer straightforward to analyse when the eigenvalues develop a non-zero real part. Specifically, it is no longer easy to see how the Krein condition from \cite{DT15} enters.

\section{Computational Analysis}\label{sec:numerics}

To illustrate the concepts we described above, we use numerical solvers implemented in MATLAB. The coefficients in the series expansions for solutions to KdV can be found analytically at all orders, but in practice it is more convenient to use floating point calculations of these coefficients. We solve \eqref{eq:zeroInt} for $p=1$ and $p=2$ for both non-resonant and resonant cases, varying the coefficient $\beta$ to get to the different regimes.

\begin{figure}[htb!]
\includegraphics[width=0.49\textwidth]{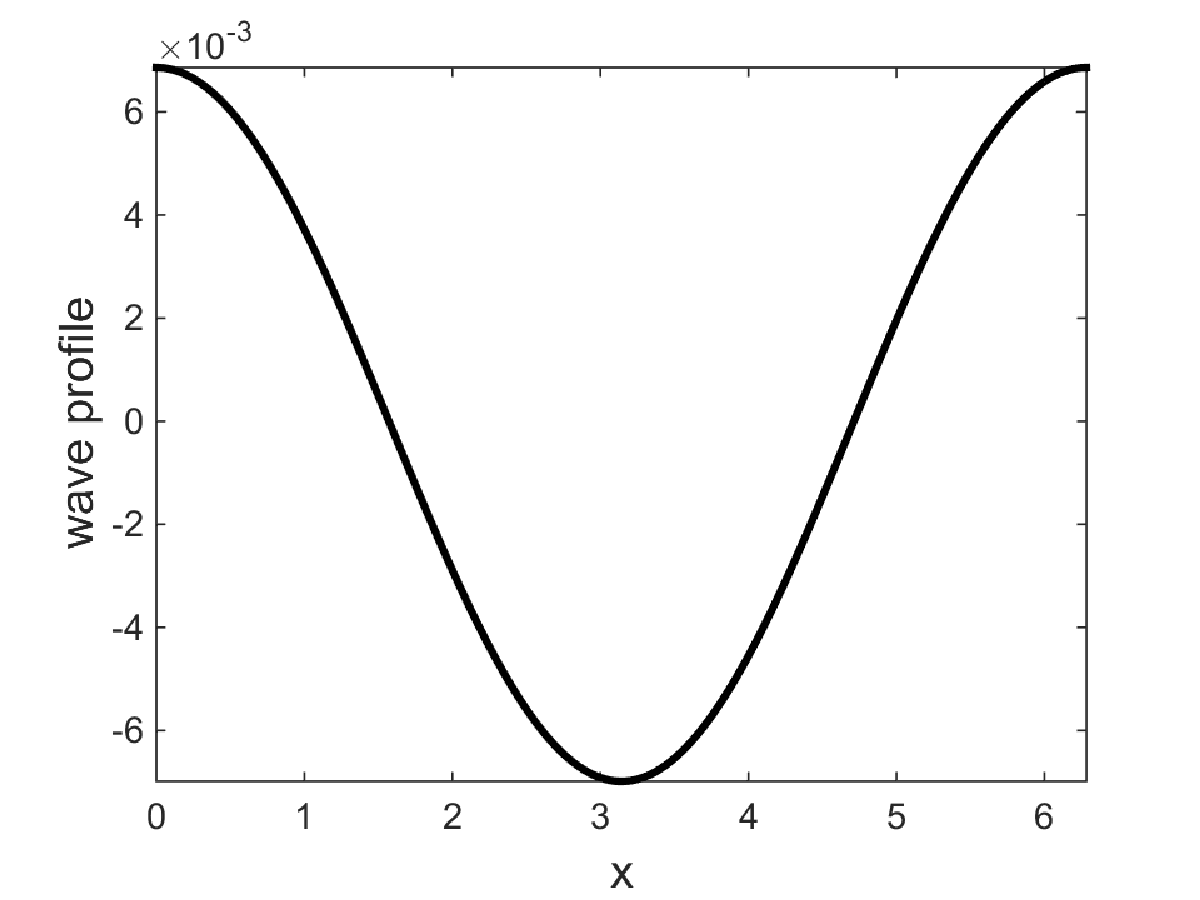}
\includegraphics[width=0.49\textwidth]{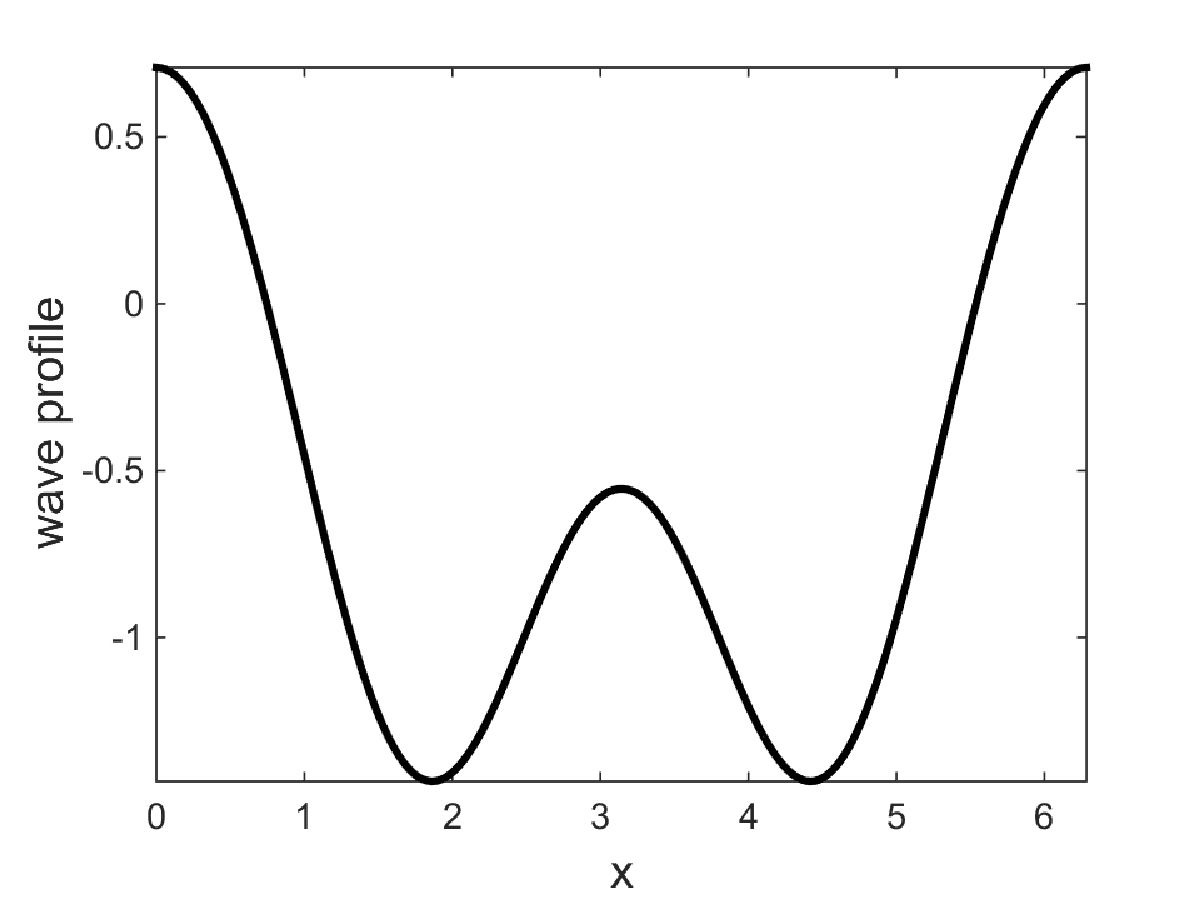}
\caption{Two sample wave profiles for \eqref{eq:travSteadyState} with $p=1$, with the wave profile for $\alpha = 1$, $\beta = 1/4$ on the left and $\alpha =1$, $\beta=1/5$ (resonant, $K=2$), on the right.\label{fig:kawaharaProfiles}}
\end{figure}

\begin{figure}[htb]
\includegraphics[width=0.49\textwidth]{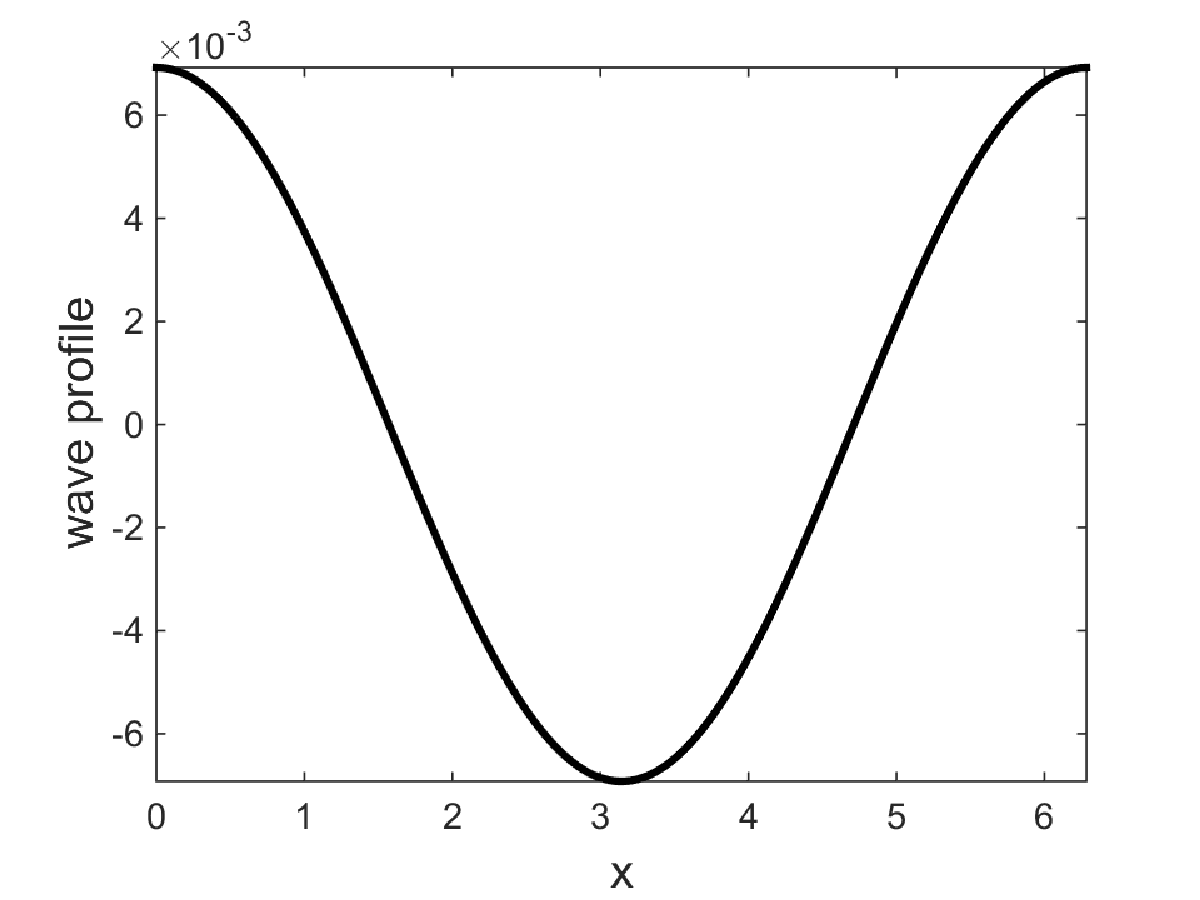}
\includegraphics[width=0.49\textwidth]{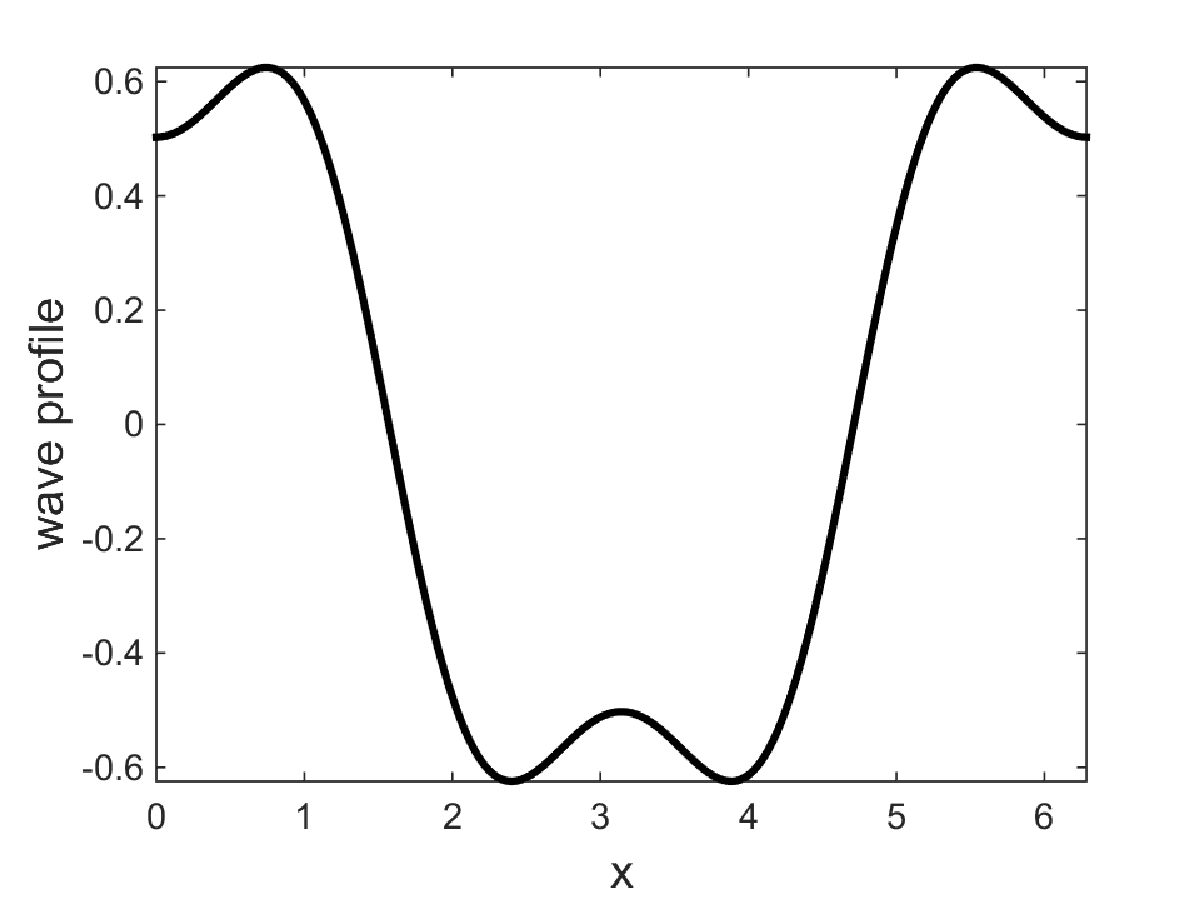}
\caption{Two sample wave profiles for \eqref{eq:cubicSteadyState} ($p=2$) with the wave profile for $\alpha = 1$, $\beta = 1/4$ on the left and $\alpha =1$, $\beta=1/10$ (resonant, $K=3$) regime, on the right. \label{fig:mKdVProfiles}}
\end{figure}

\begin{figure}[htb]
\includegraphics[width=0.32\textwidth]{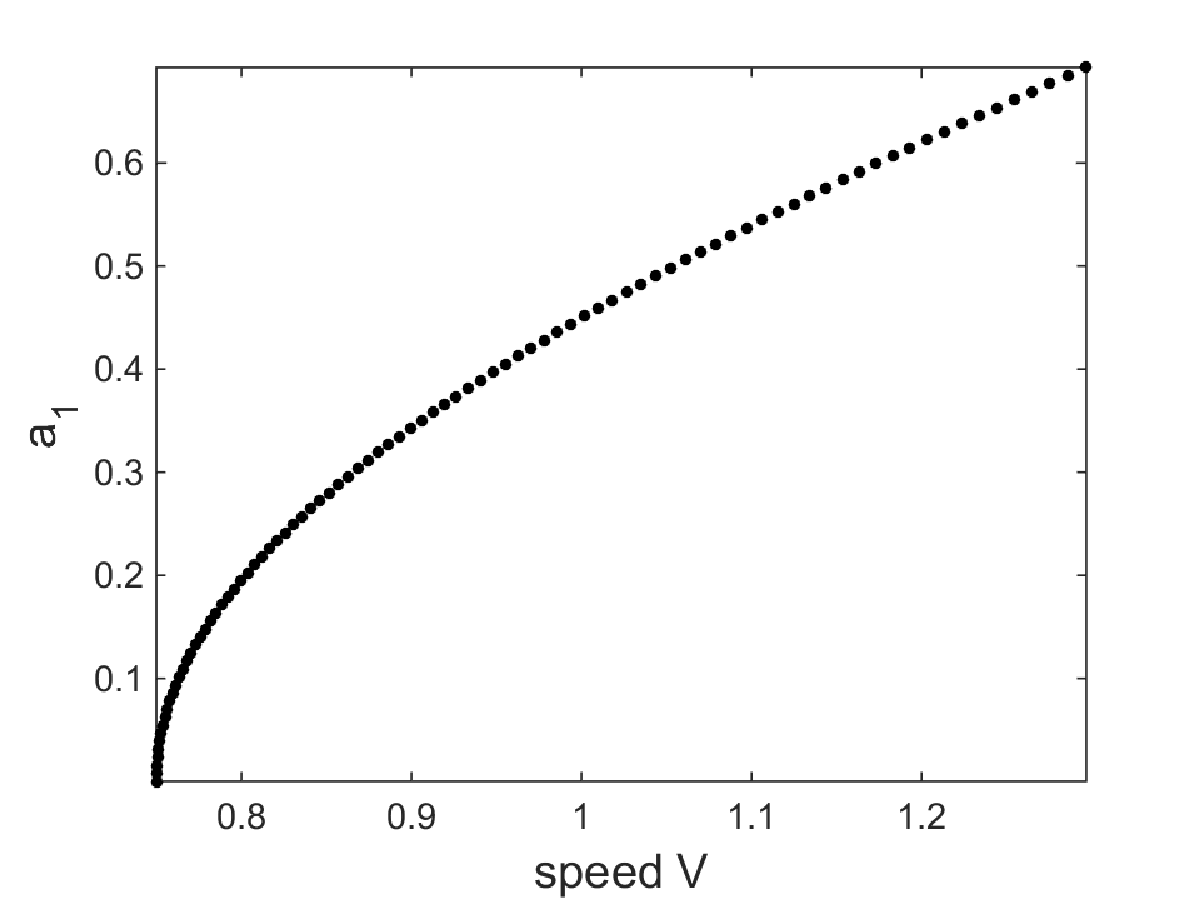}
\includegraphics[width=0.32\textwidth]{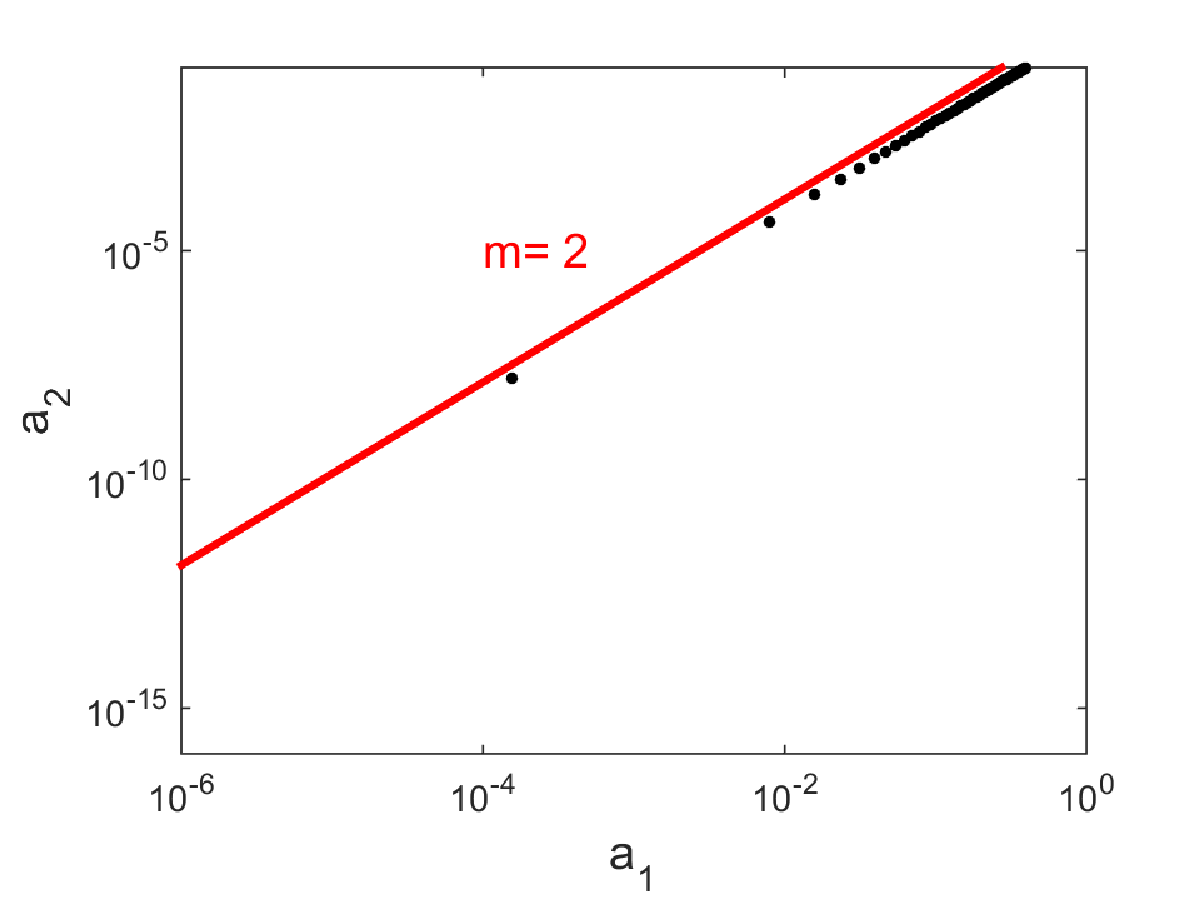}
\includegraphics[width=0.32\textwidth]{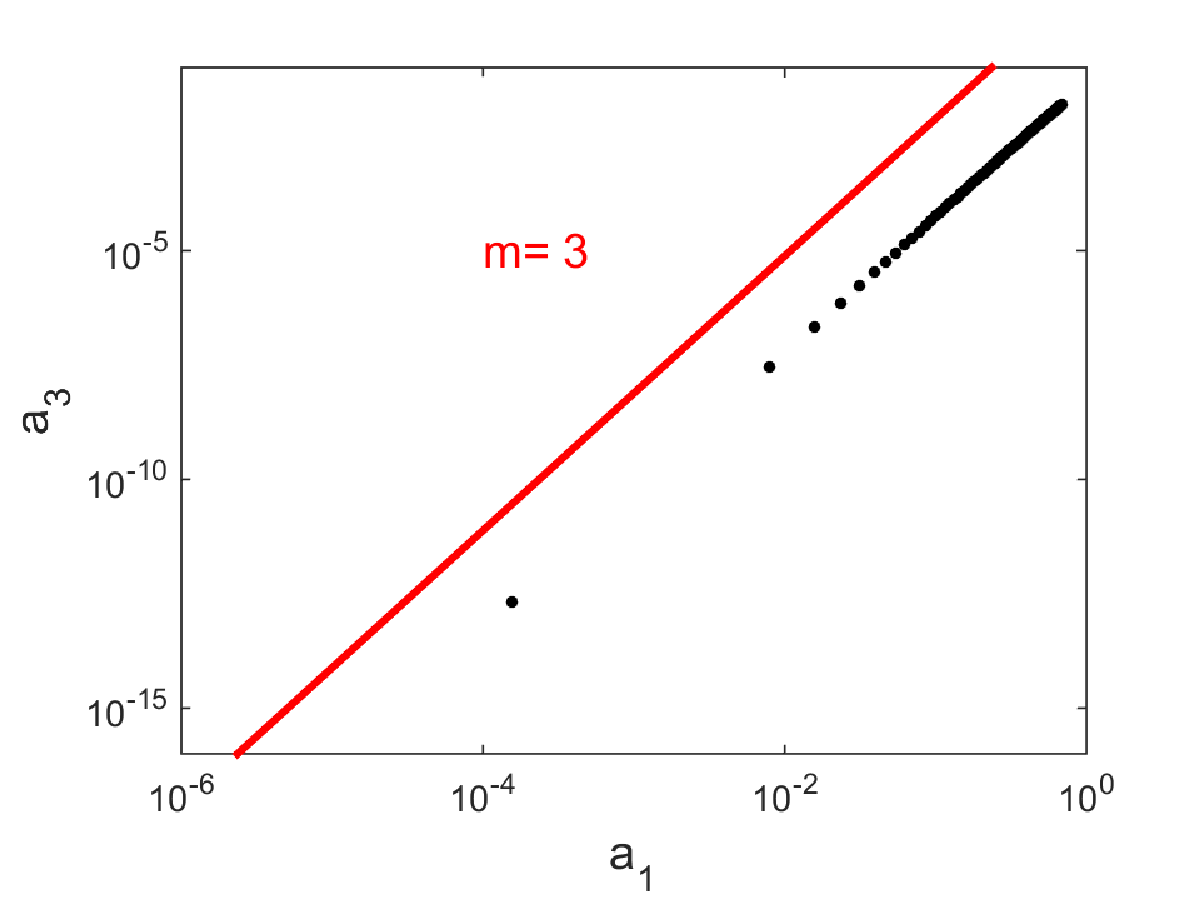}\\
\includegraphics[width=0.32\textwidth]{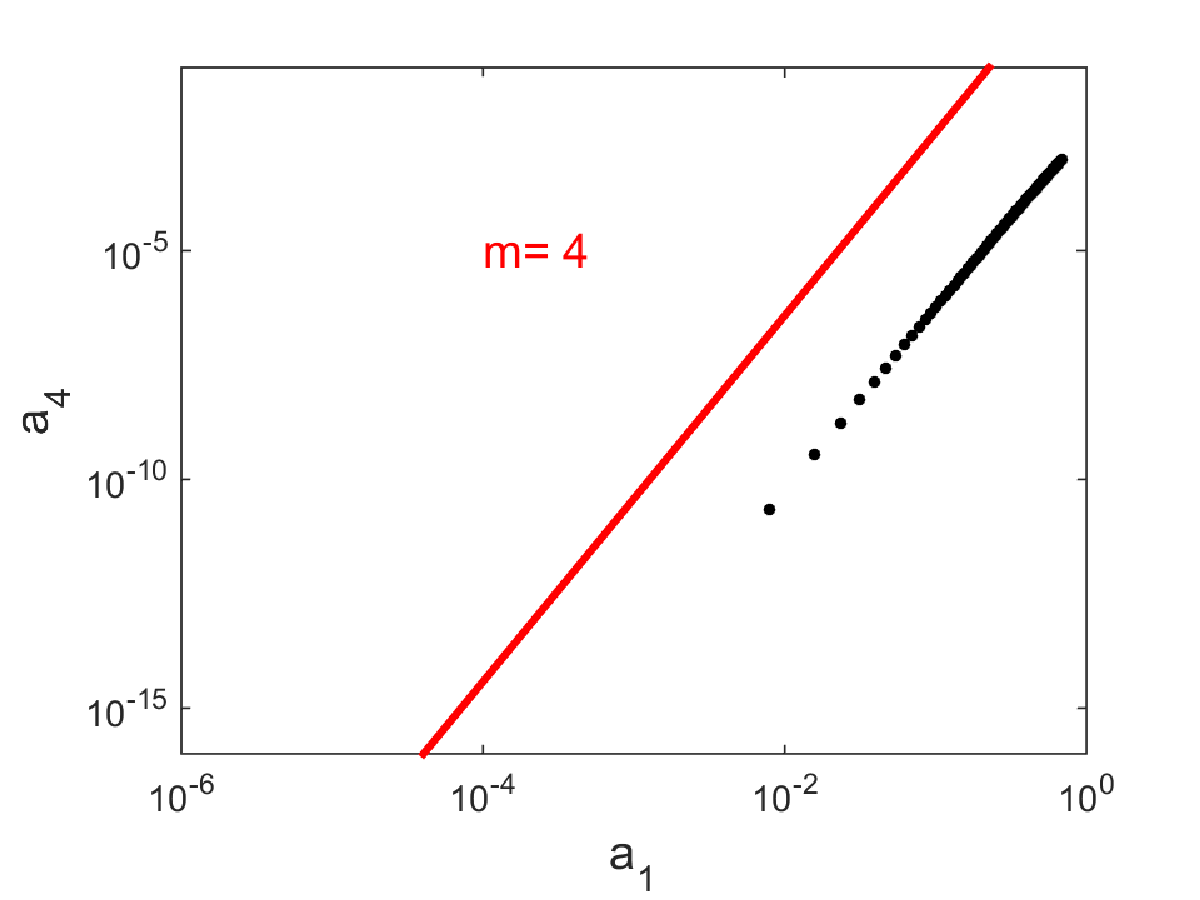}
\includegraphics[width=0.32\textwidth]{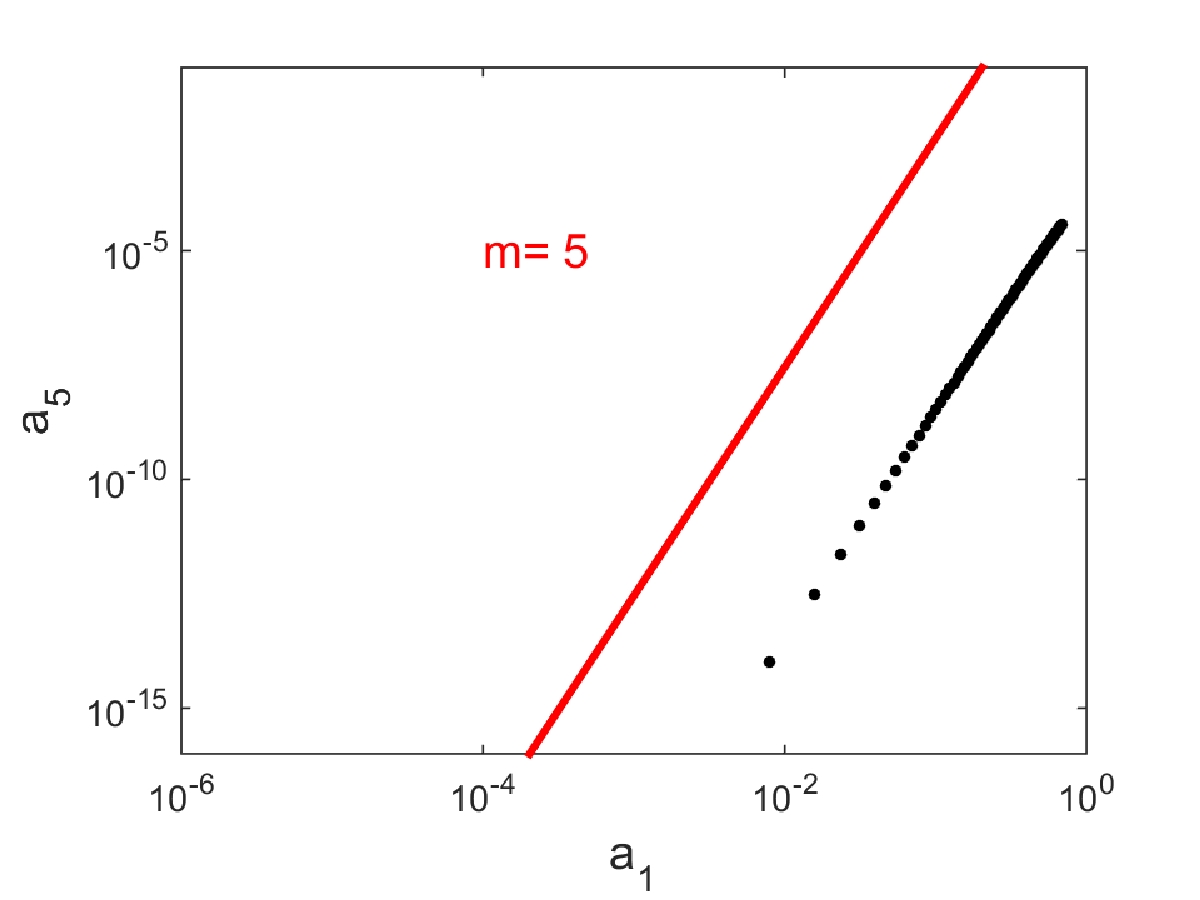}
\includegraphics[width=0.32\textwidth]{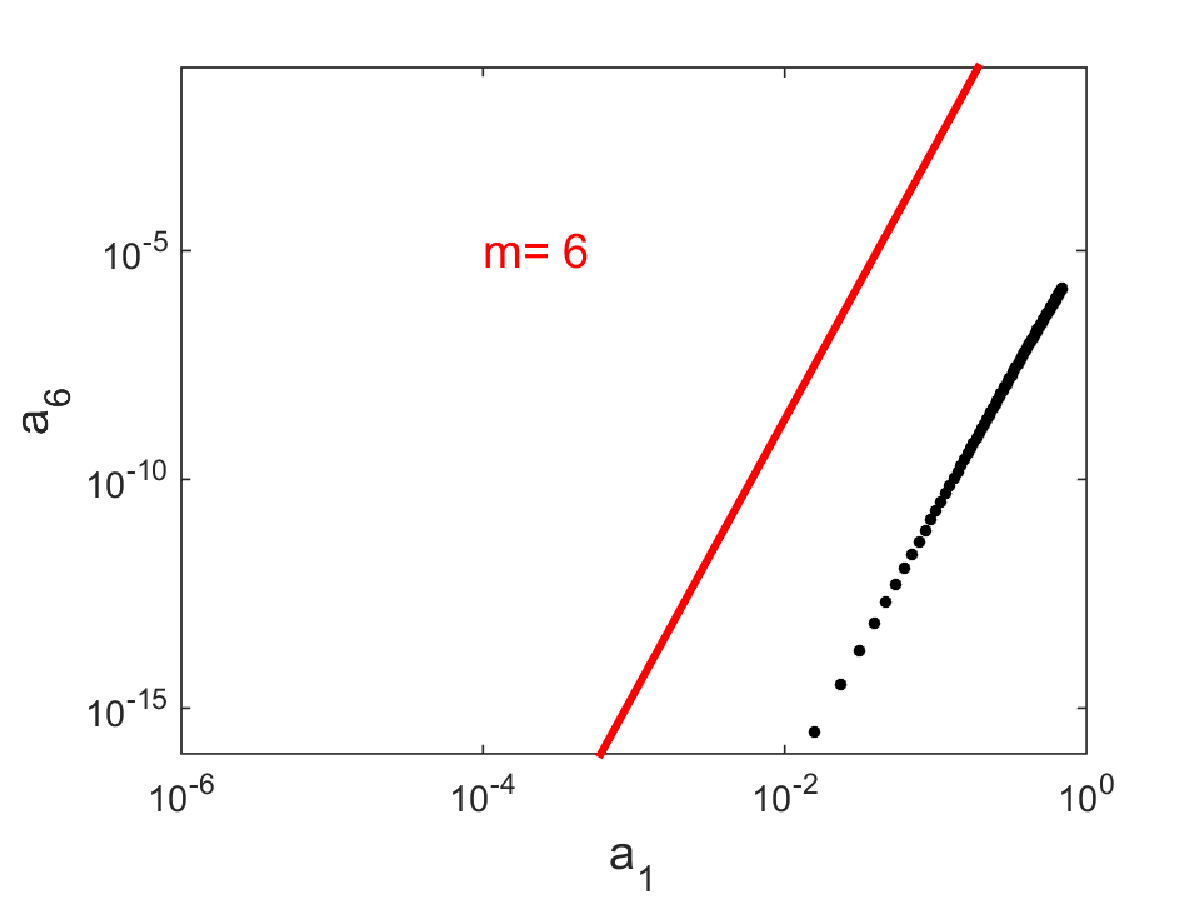}
\caption{Fourier oefficients of a branch of solutions of \eqref{eq:travSteadyState} with $p=1$, with $\alpha = 1$ and $\beta = 1/4$.	The top left figure shows the bifurcation branch of the first Fourier mode $a_1$ as a function of the speed of the wave $V$. Subsequent plots display Fourier coefficients $a_j$ vs. $a_1$ (i.e., the small parameter) on a log-log plot compared to a line with the labelled slope. We see that $a_k=O(\epsilon^k)$.
 \label{fig:kawaModesbeta1_4}}
\end{figure}

\begin{figure}[htb]
\includegraphics[width=0.32\textwidth]{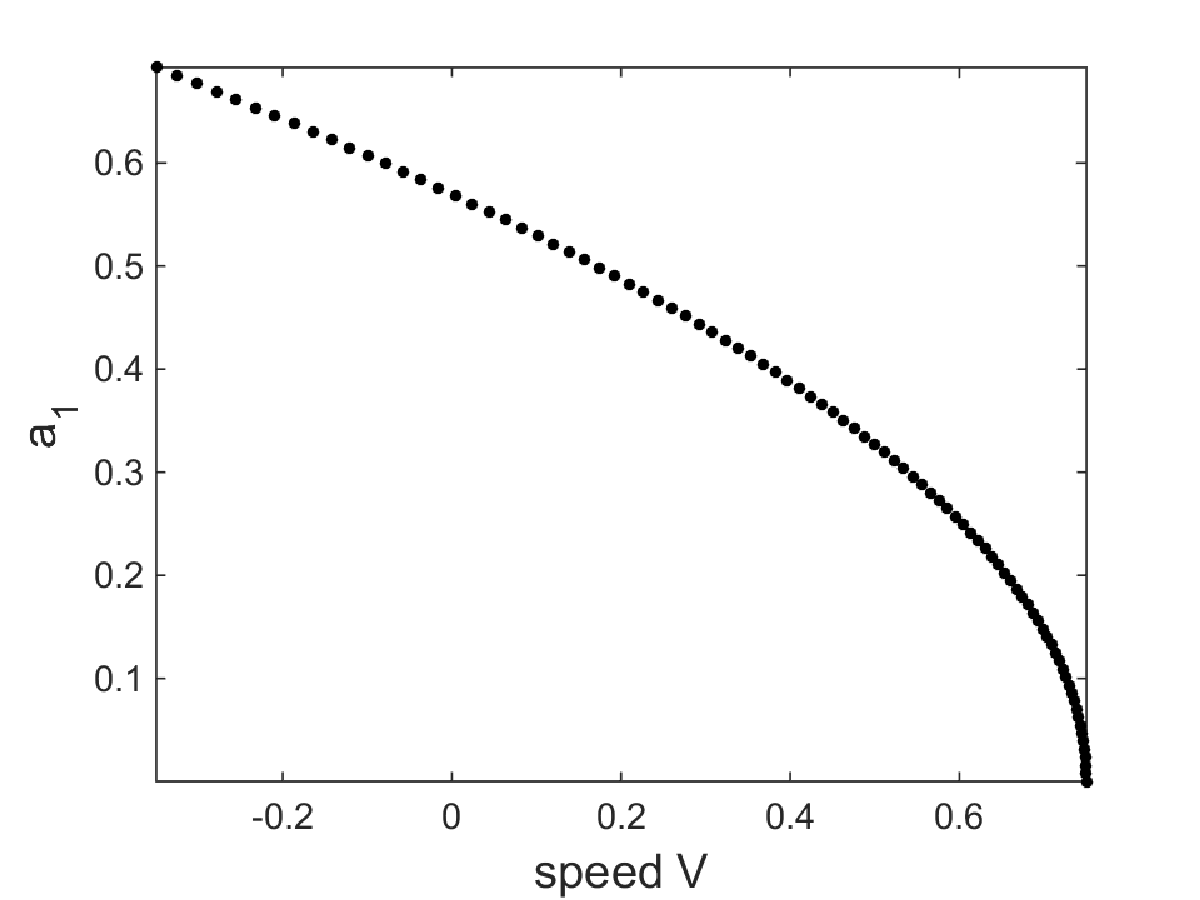}
\includegraphics[width=0.32\textwidth]{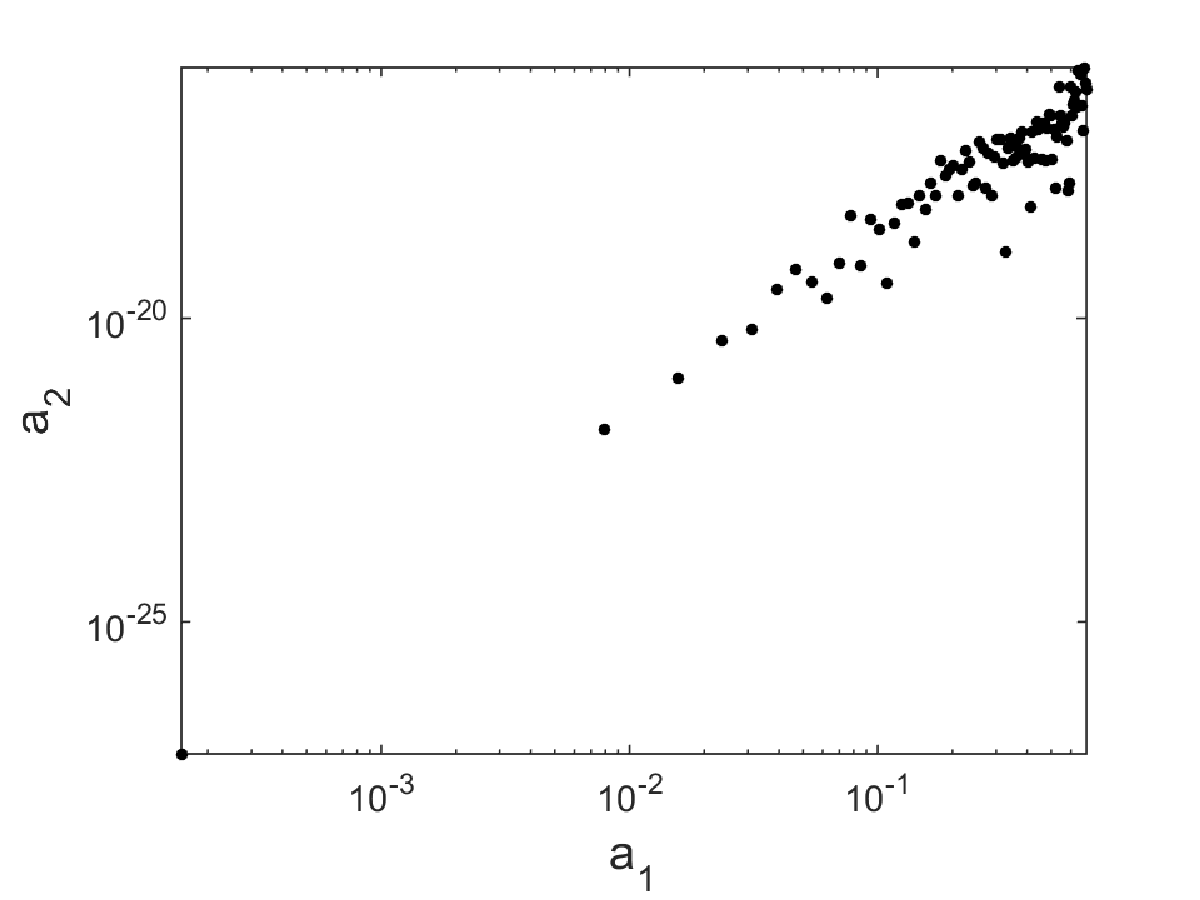}
\includegraphics[width=0.32\textwidth]{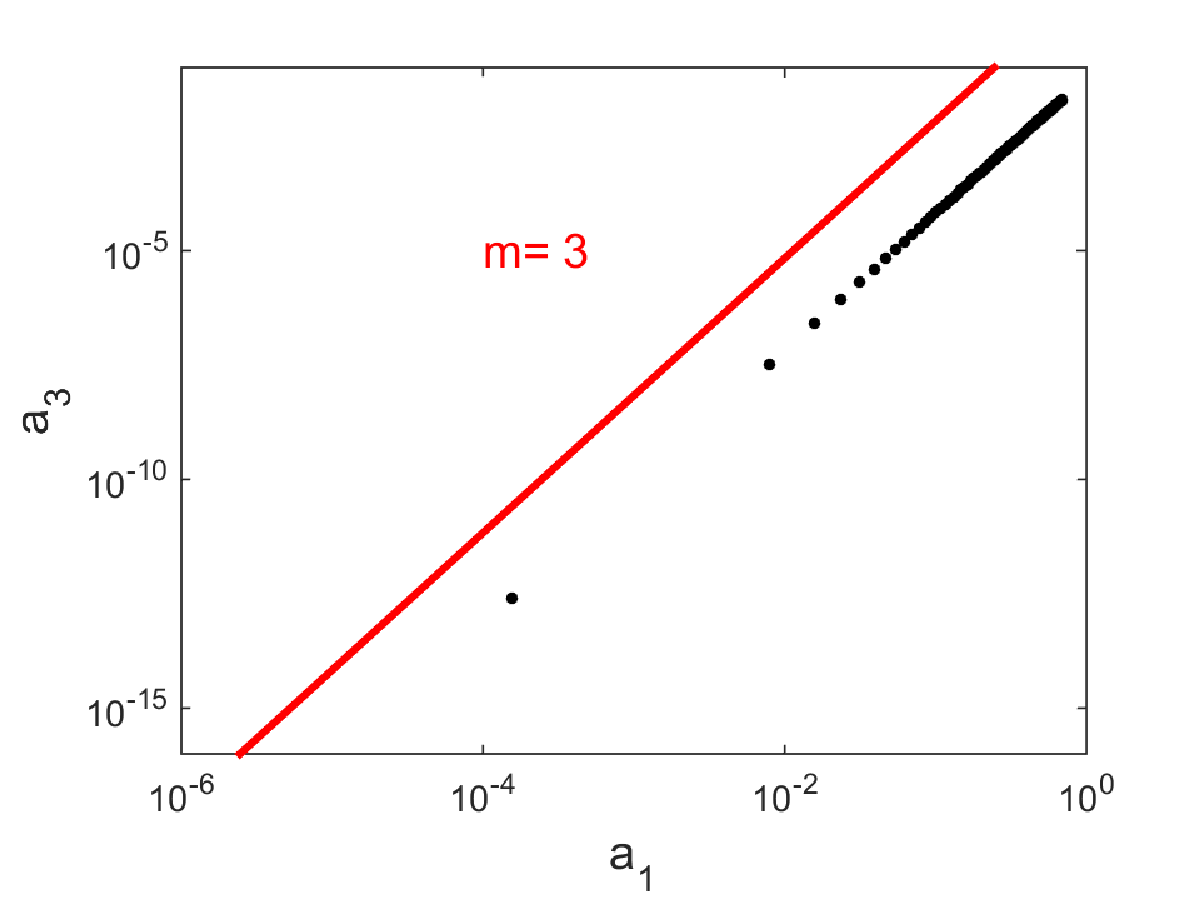}\\
\includegraphics[width=0.32\textwidth]{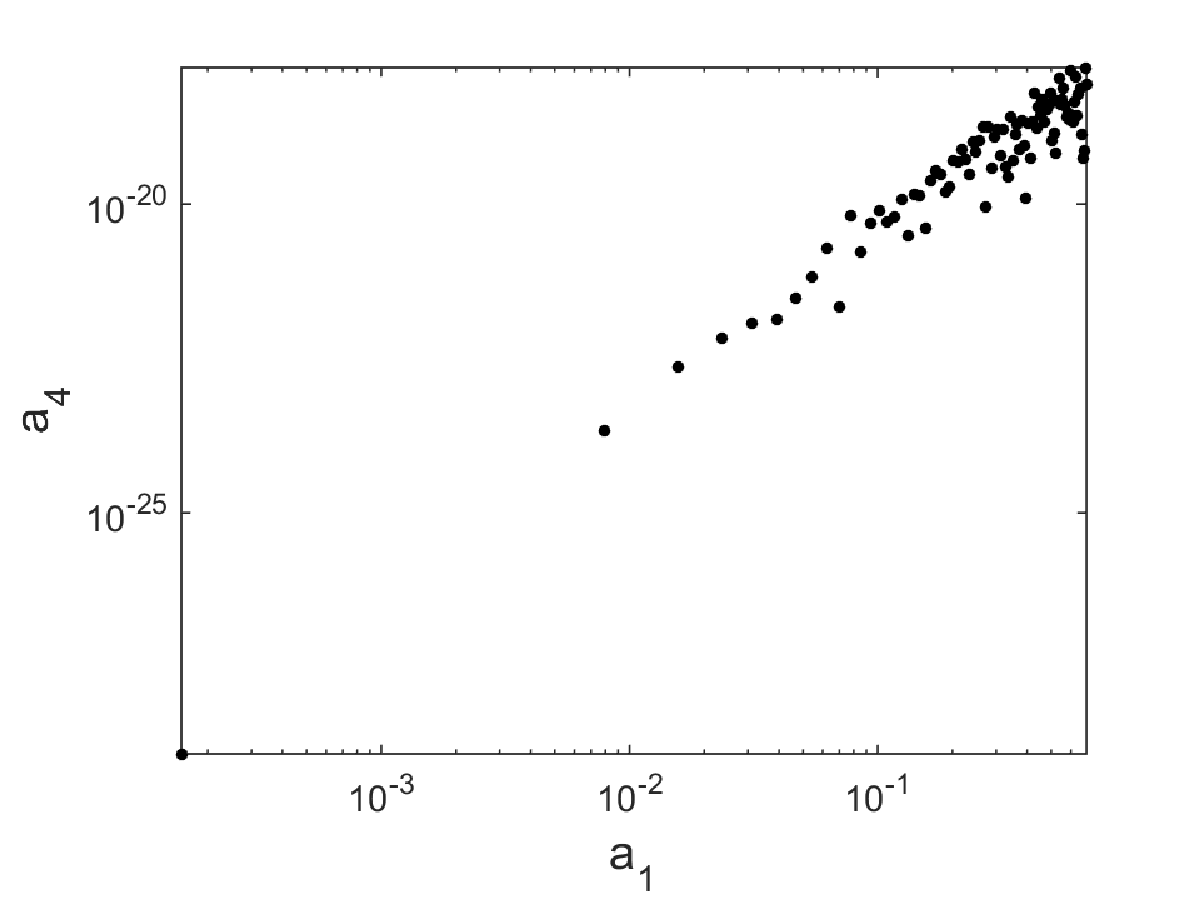}
\includegraphics[width=0.32\textwidth]{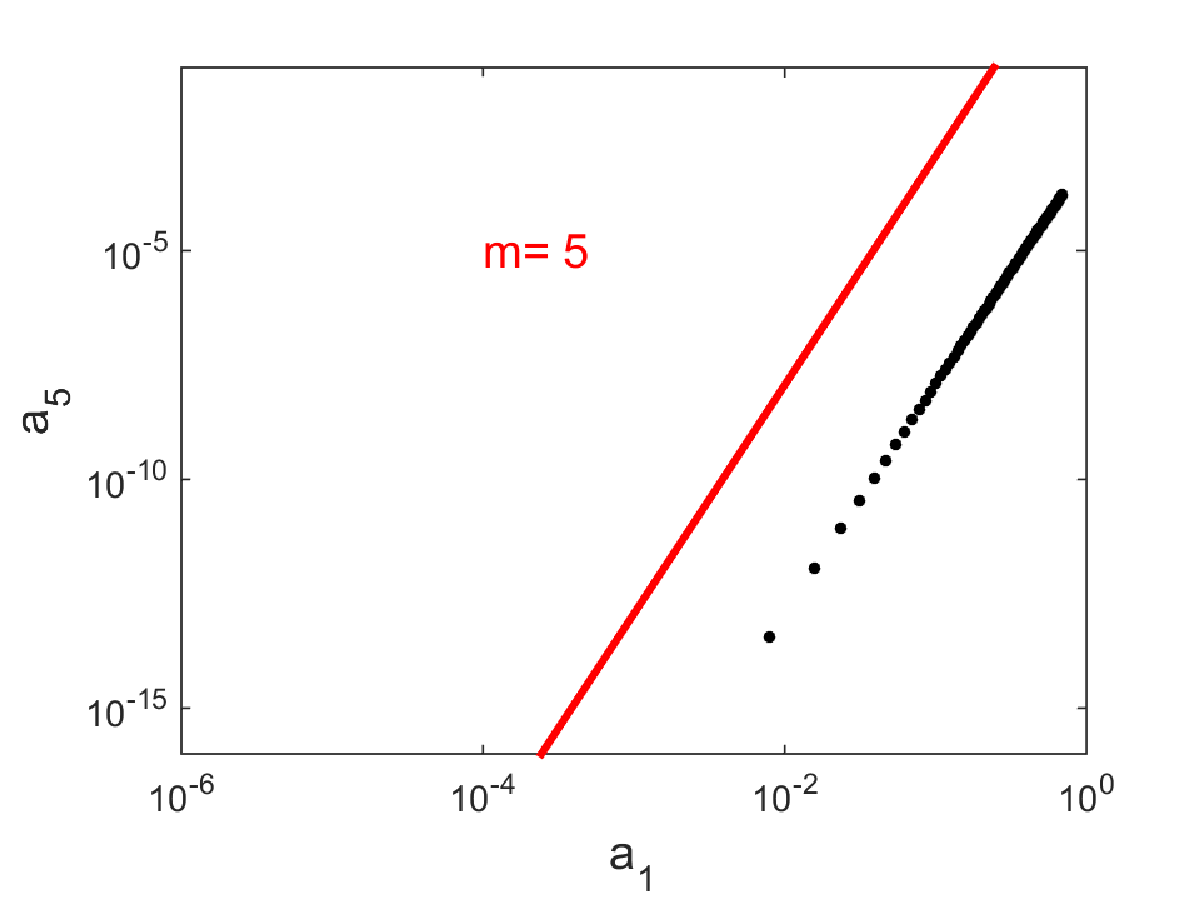}
\includegraphics[width=0.32\textwidth]{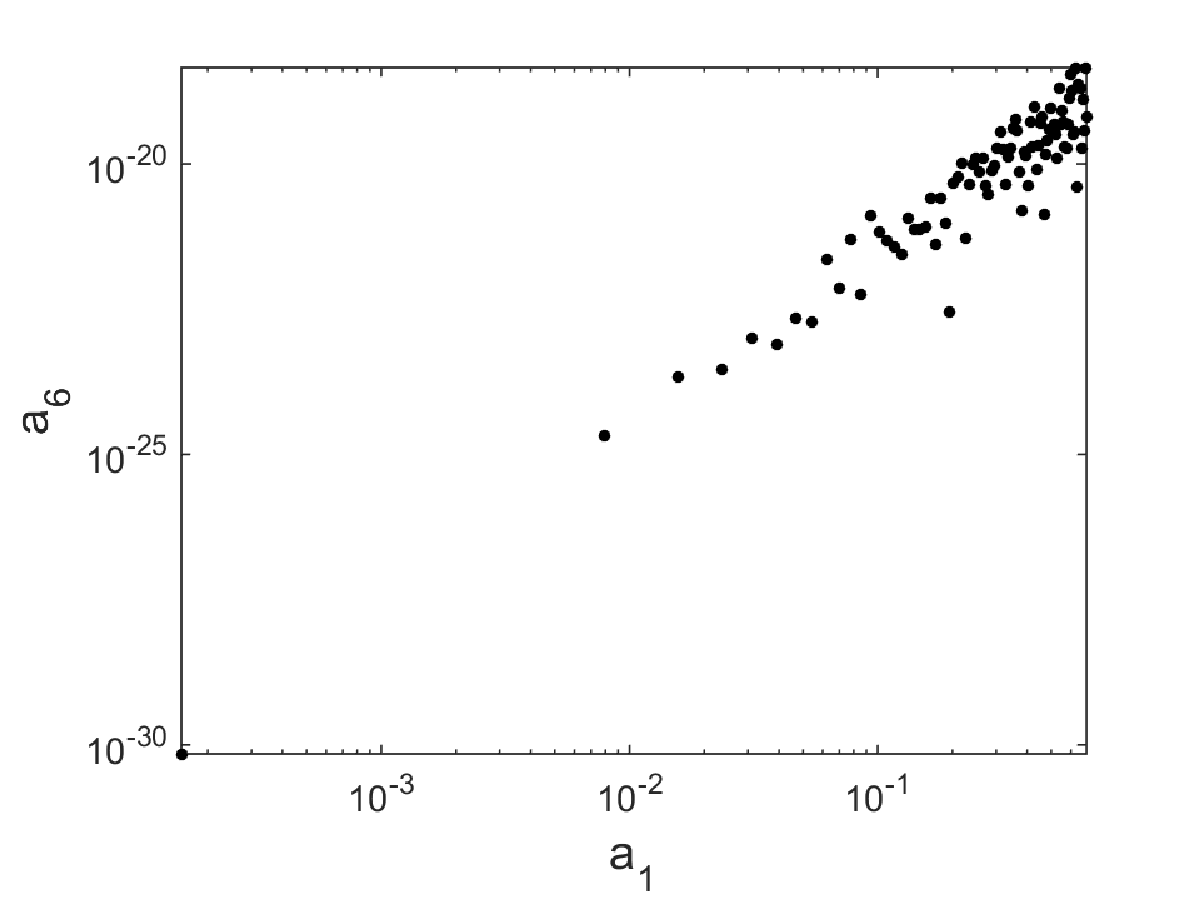}
\caption{Fourier coefficients of a branch of solutions of \eqref{eq:cubicSteadyState} ($p=2$), with $\alpha = 1$ and $\beta = 1/4$. The top left figure shows the bifurcation branch of $a_1$ as a function of $V$. Subsequent plots display Fourier coefficients $a_j$ vs. $a_1$ (i.e., the small parameter) on a log-log plot fitted to a line with the labelled slope. We see that $a_k =O(\epsilon^k)$ with even-order coefficients equal to zero.
\label{fig:mKdVModesbeta1_4}}
\end{figure}

\begin{figure}[htb]
\includegraphics[width=0.32\textwidth]{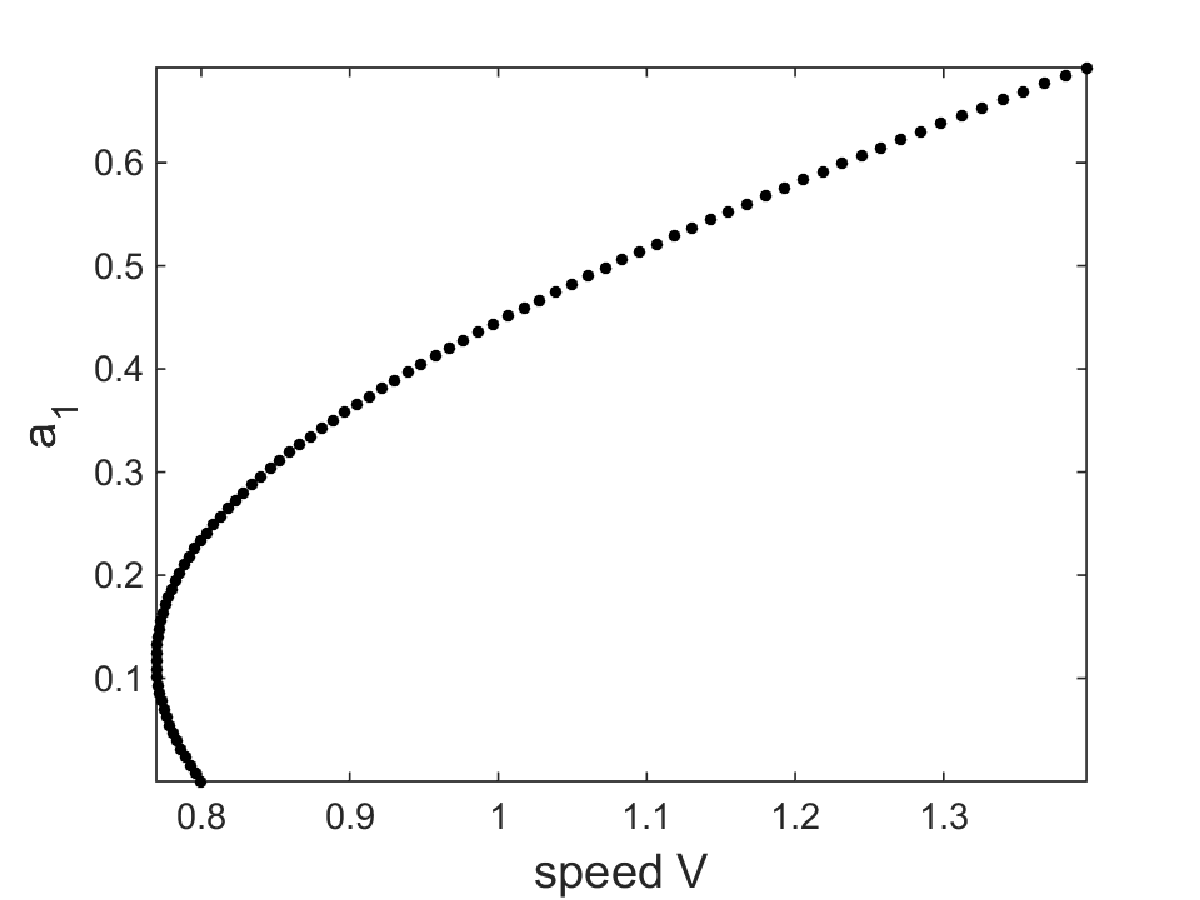}
\includegraphics[width=0.32\textwidth]{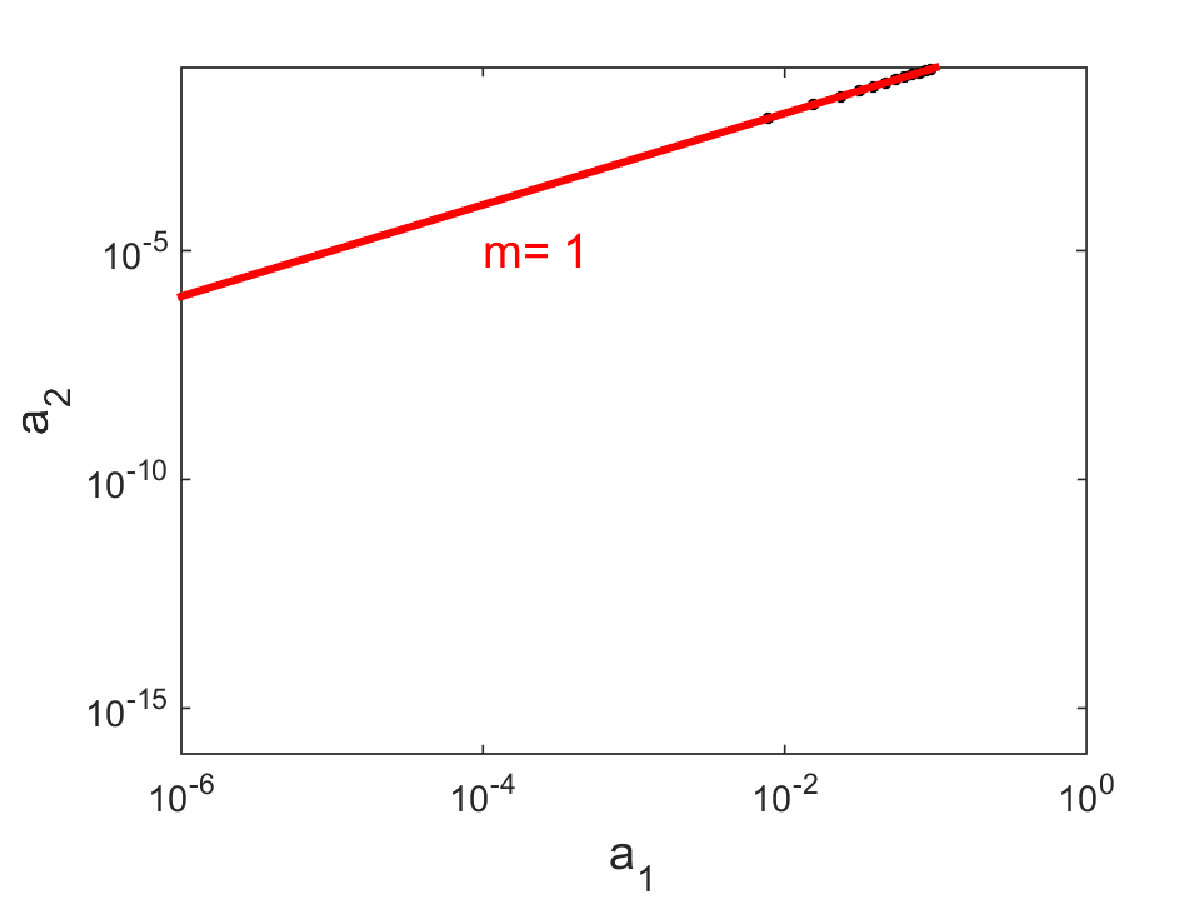}
\includegraphics[width=0.32\textwidth]{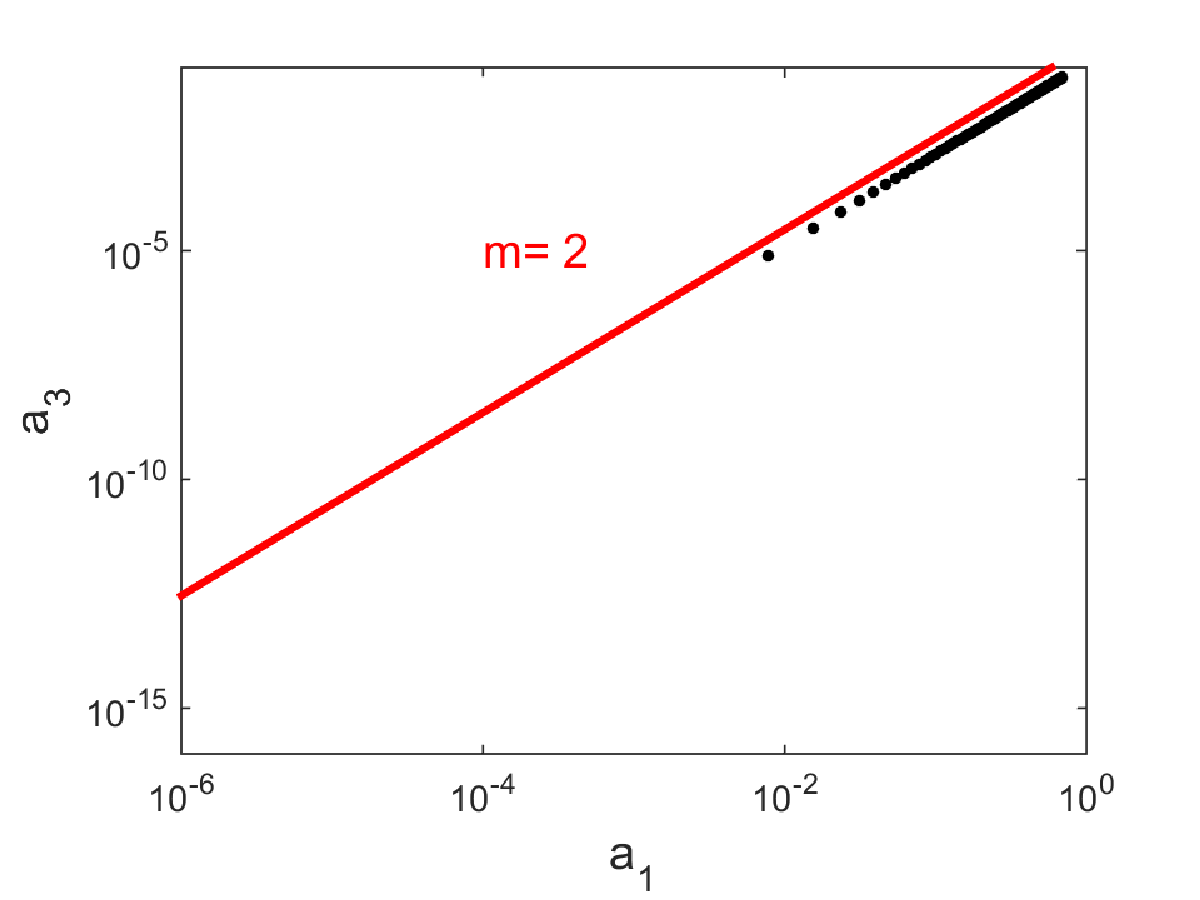}\\
\includegraphics[width=0.32\textwidth]{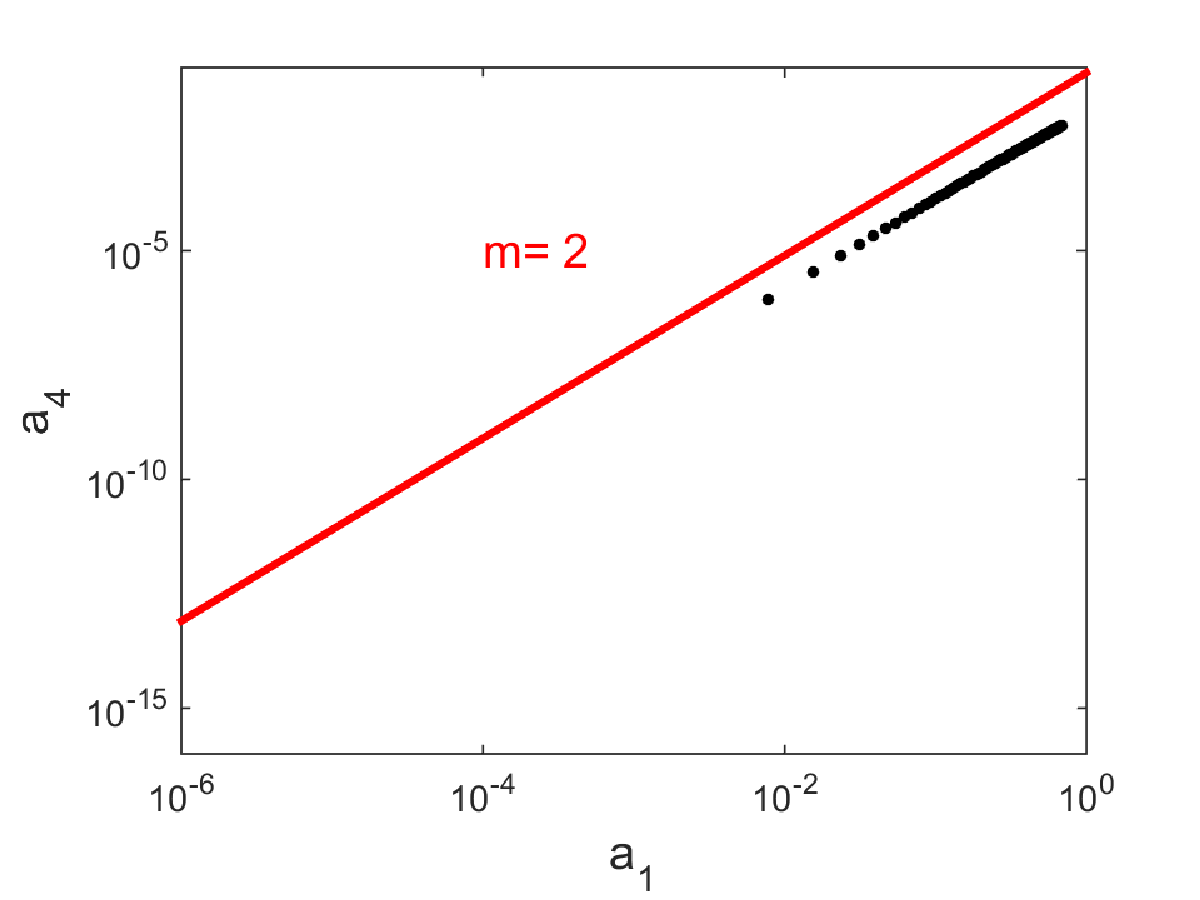}
\includegraphics[width=0.32\textwidth]{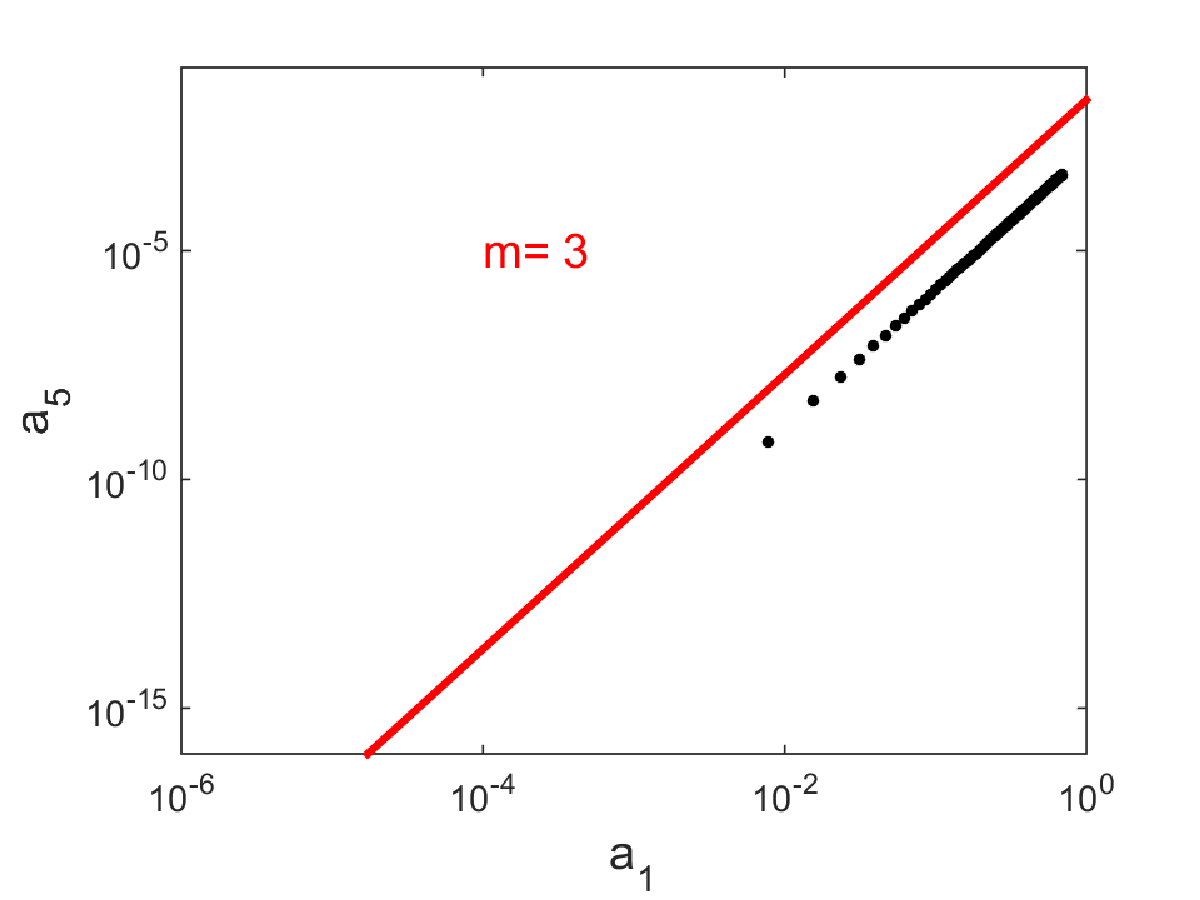}
\includegraphics[width=0.32\textwidth]{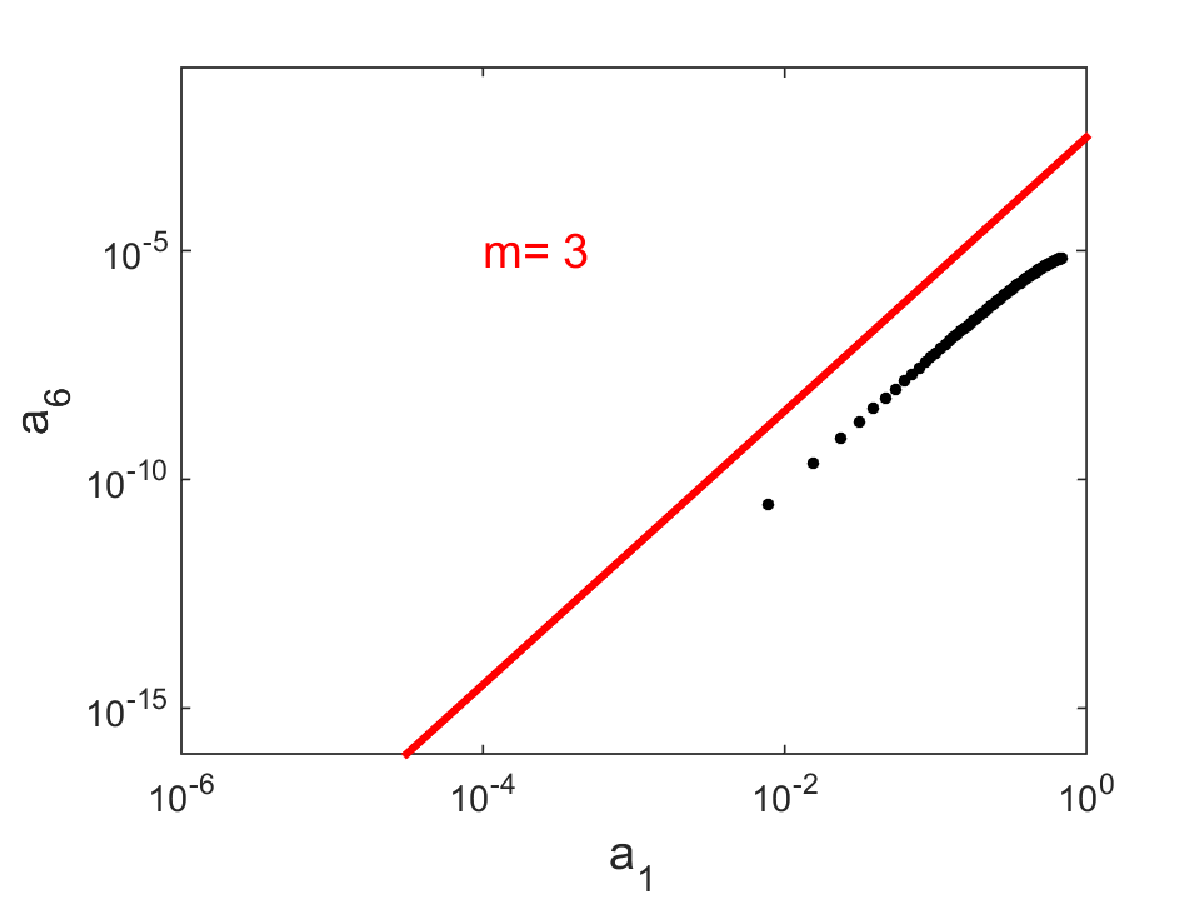}
\caption{Fourier coefficients of a branch of solutions of \eqref{eq:travSteadyState} with $p=1$, in the resonant regime with $K=2$, $\alpha = 1$ and $\beta = 1/5$.	The top left figure is the bifurcation branch of $a_1$ as a function of $V$. Subsequent plots display Fourier coefficients plotted versus the first Fourier mode $a_1$ (i.e., the small parameter) on a log-log plot compared to a line with the labelled slope. We see that both $a_1=O(\epsilon)$ (by assumption) and $a_2=O(\epsilon)$. \label{fig:kawaModesbeta1_5}}
\end{figure}

\begin{figure}[htb]
\includegraphics[width=0.32\textwidth]{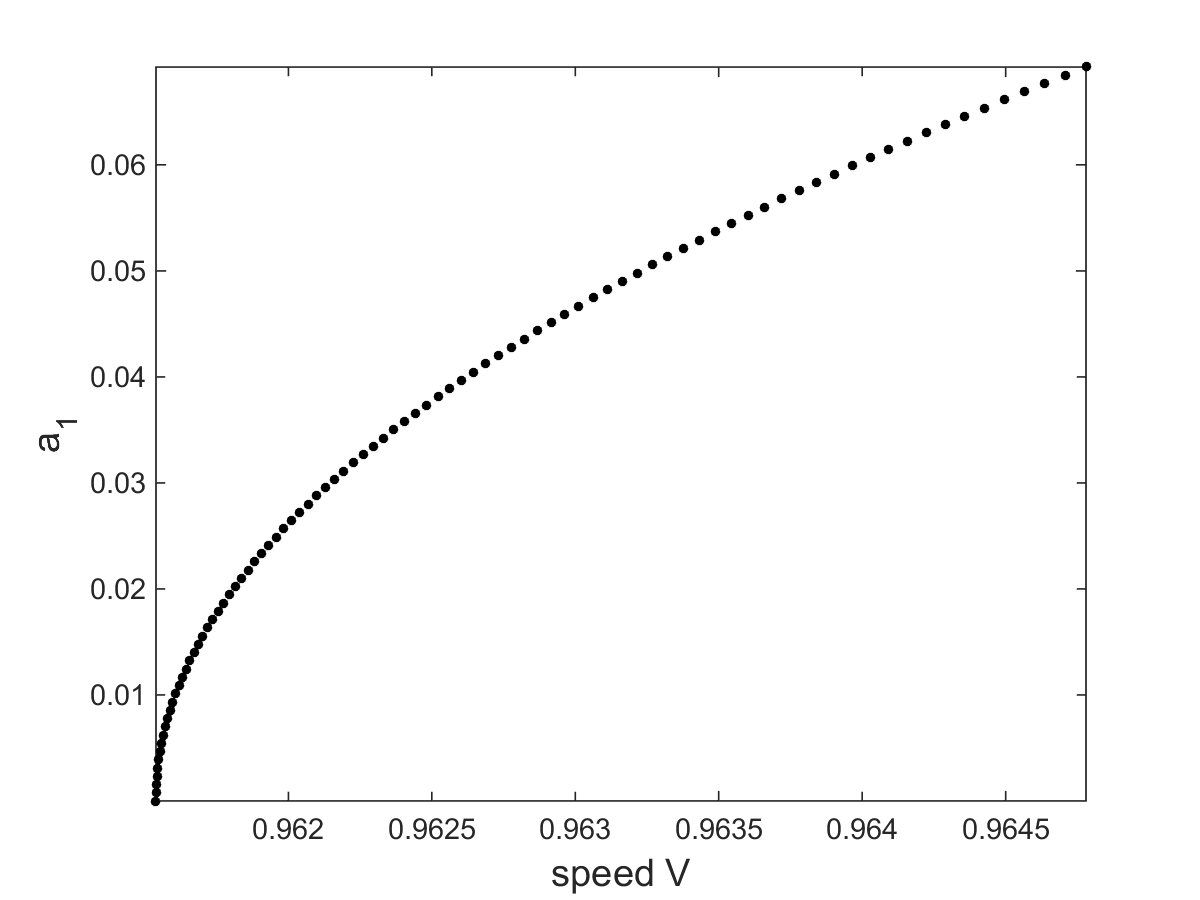}
\includegraphics[width=0.32\textwidth]{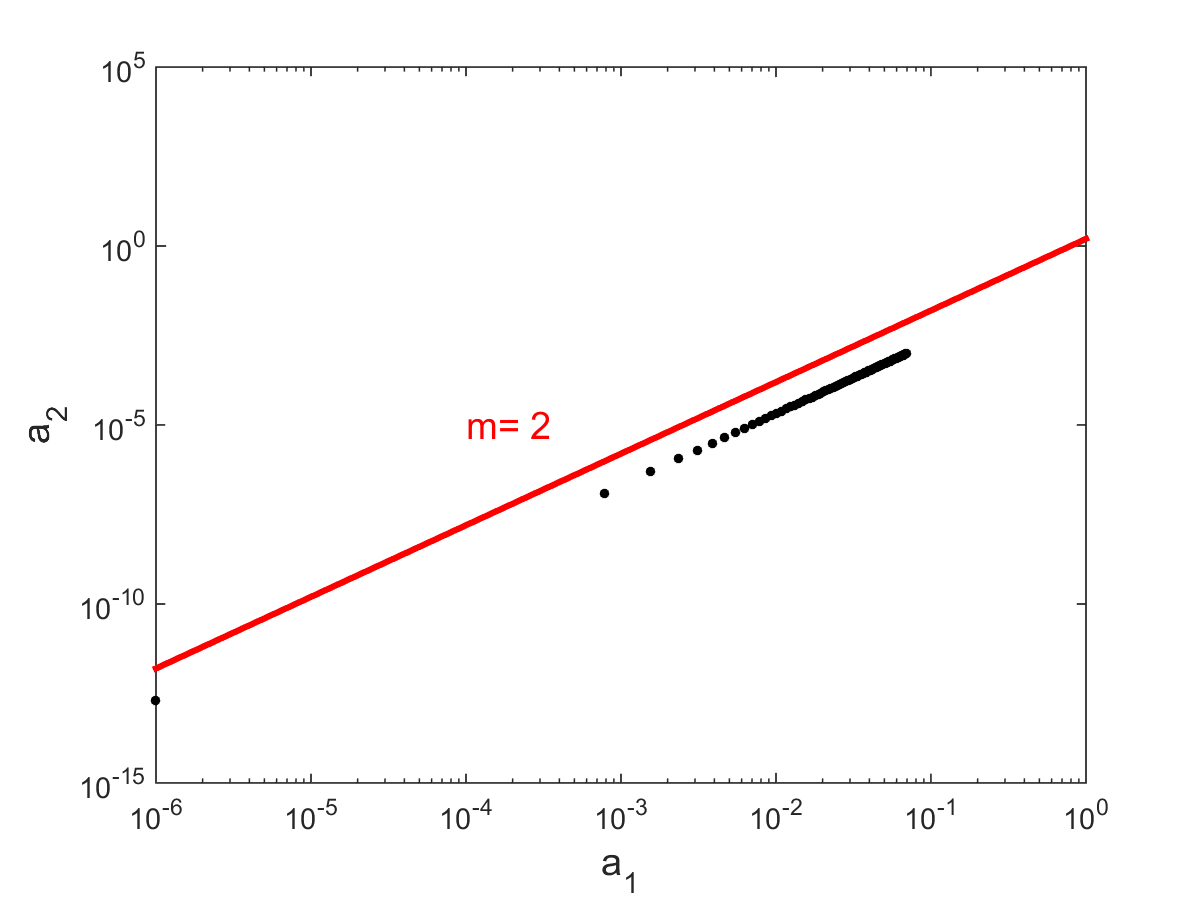}
\includegraphics[width=0.32\textwidth]{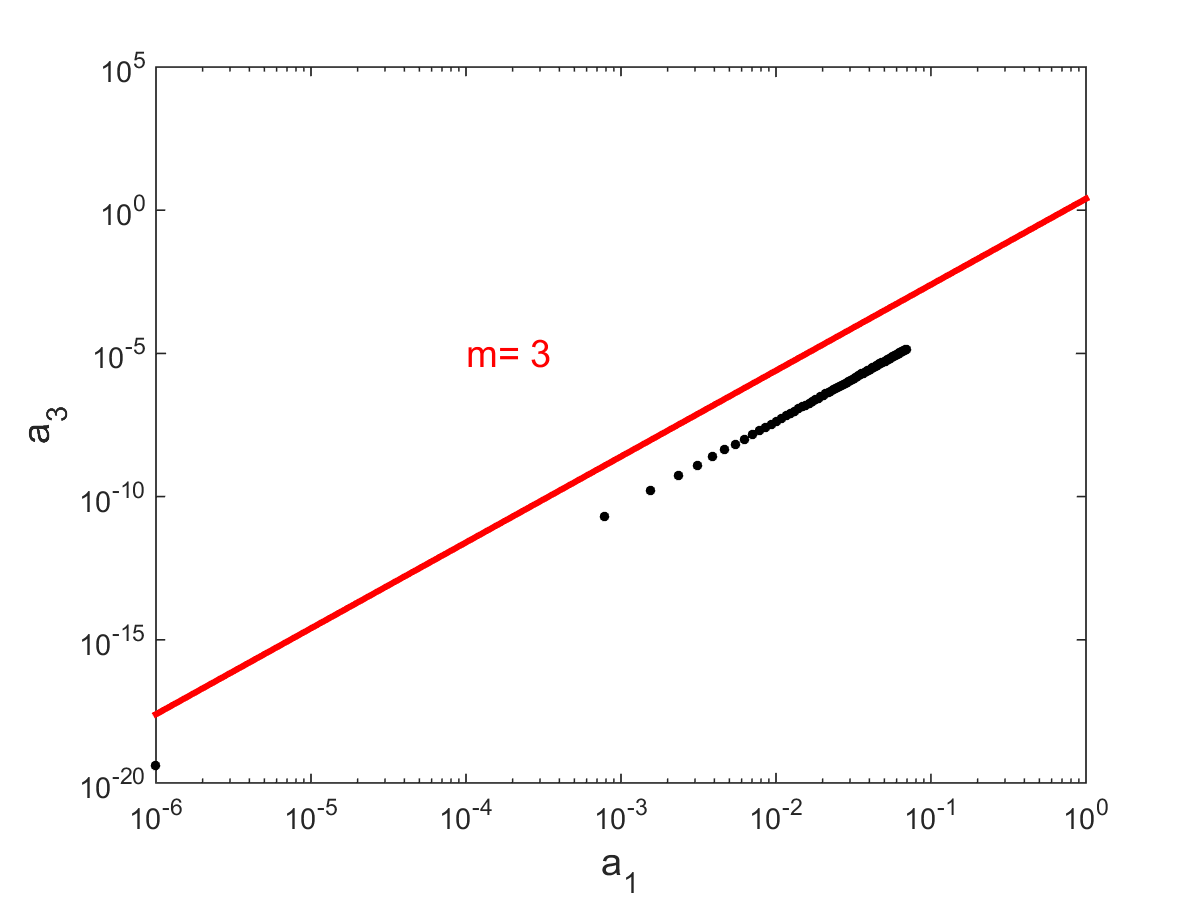}\\
\includegraphics[width=0.32\textwidth]{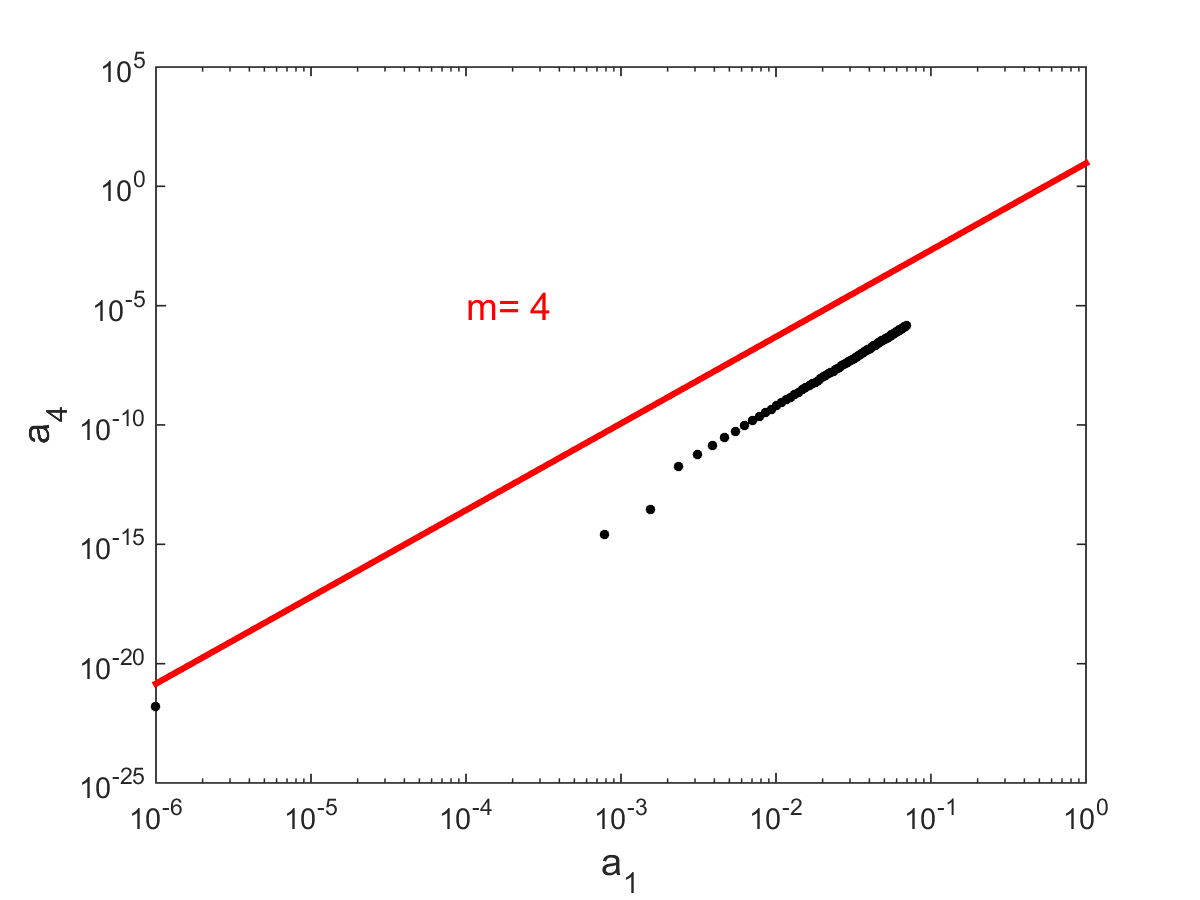}
\includegraphics[width=0.32\textwidth]{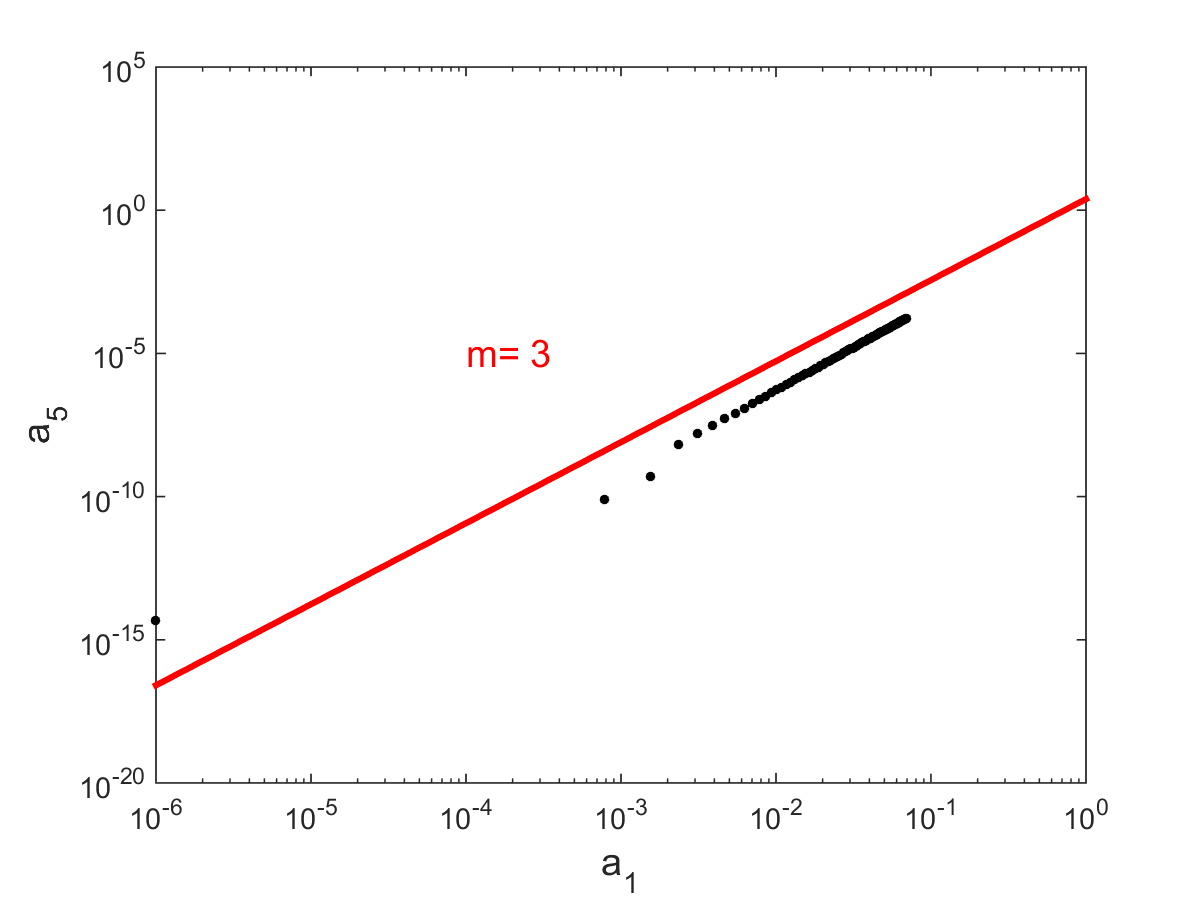}
\includegraphics[width=0.32\textwidth]{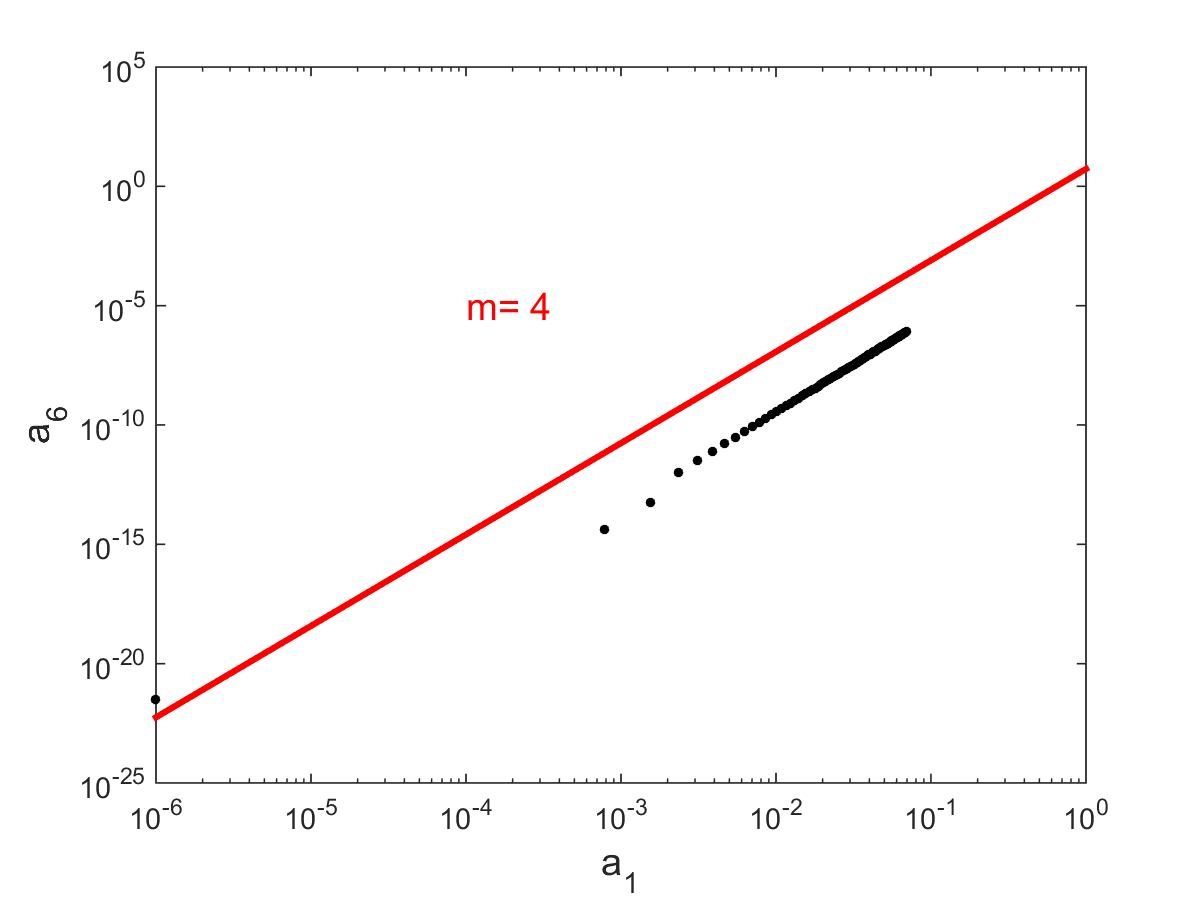}
\caption{Fourier coefficients of a branch of solutions of  \eqref{eq:travSteadyState} with $p=1$ in the resonant regime with $K=5$, $\alpha = 1$ and $\beta = 1/26$. The top left figure shows the bifurcation branch of $a_1$ as it depends on $V$, followed by log-log plots of $a_j$ vs. $a_1$, and a line with the labelled slope. We have that $a_1 = O(\epsilon)$ (by assumption) and the resonant mode $a_5 = O(\epsilon^{3})$. \label{fig:kawaModesbeta1_26}}
\end{figure}

\begin{figure}[htb]
\includegraphics[width=0.32\textwidth]{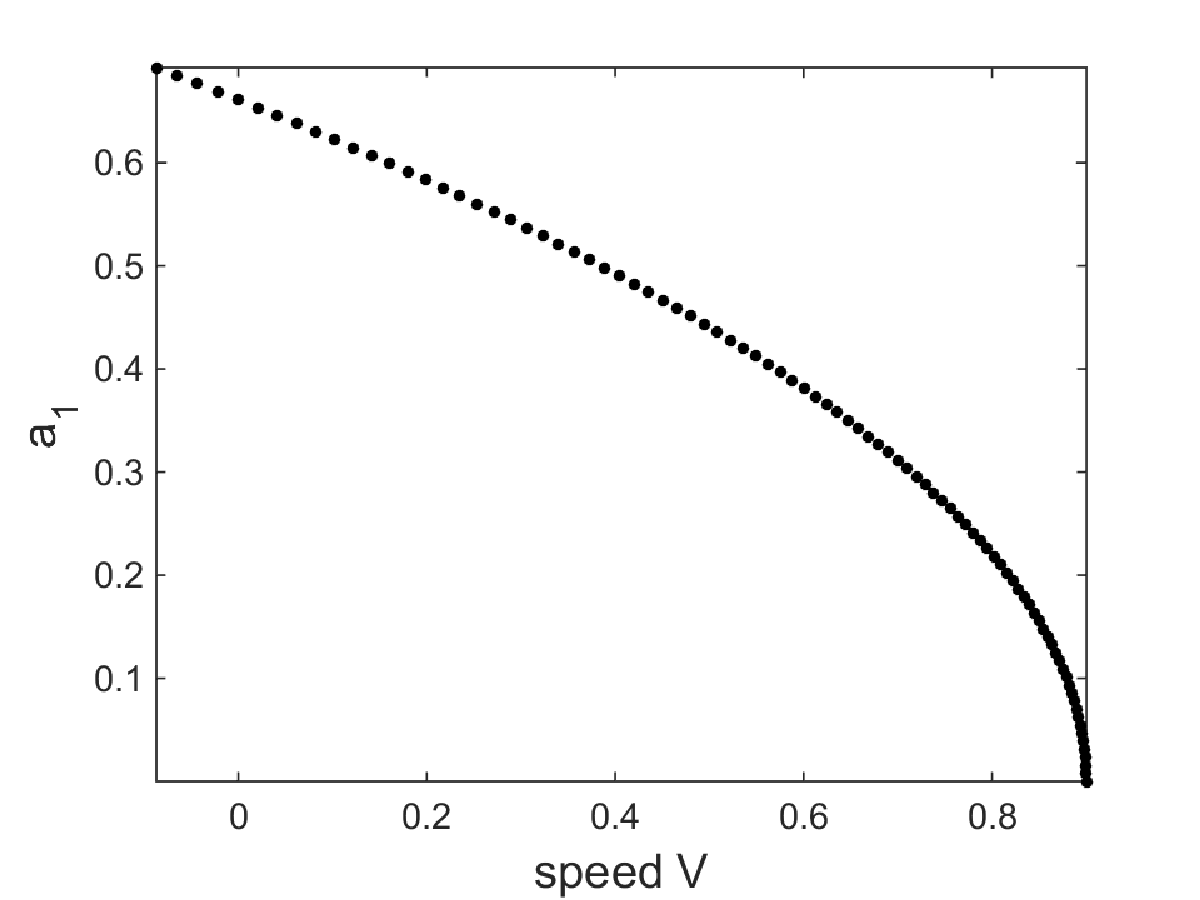}
\includegraphics[width=0.32\textwidth]{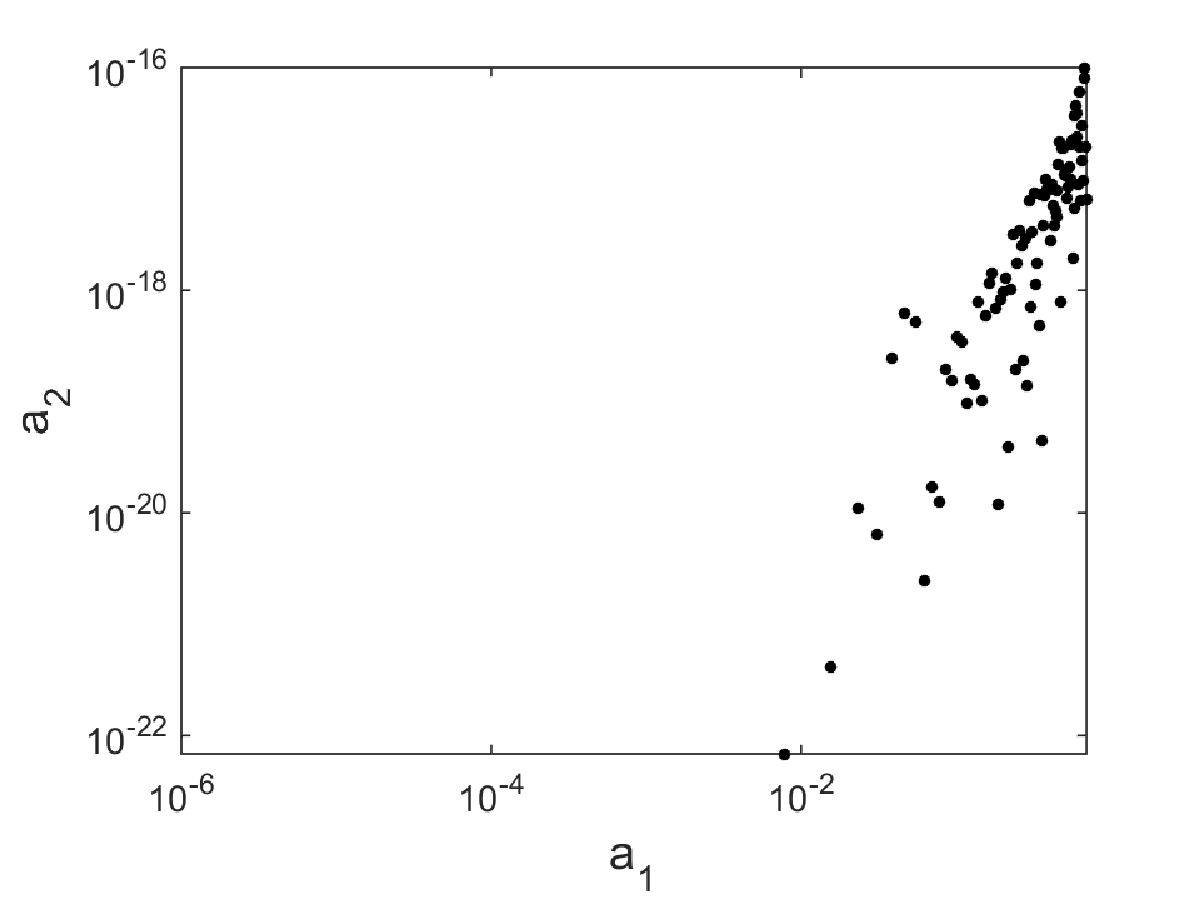}
\includegraphics[width=0.32\textwidth]{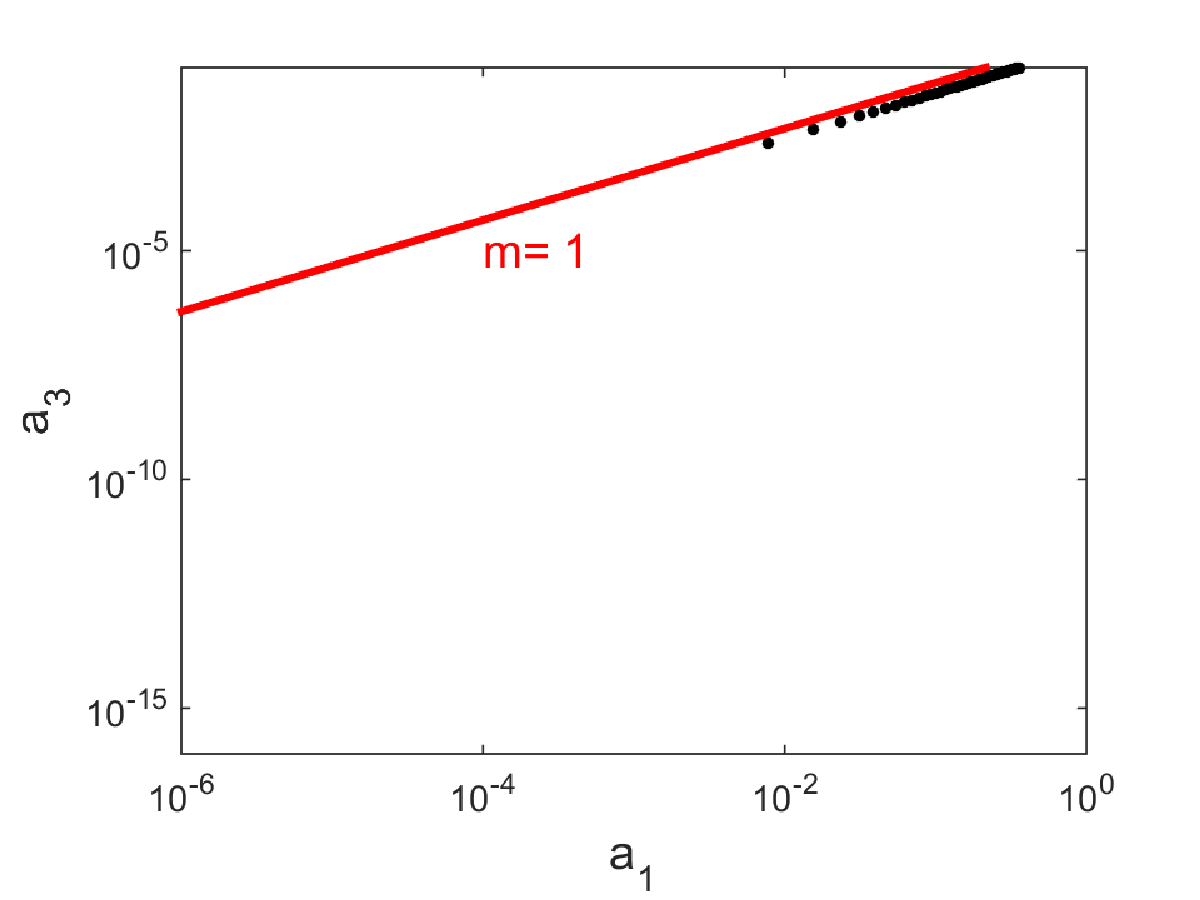}\\
\includegraphics[width=0.32\textwidth]{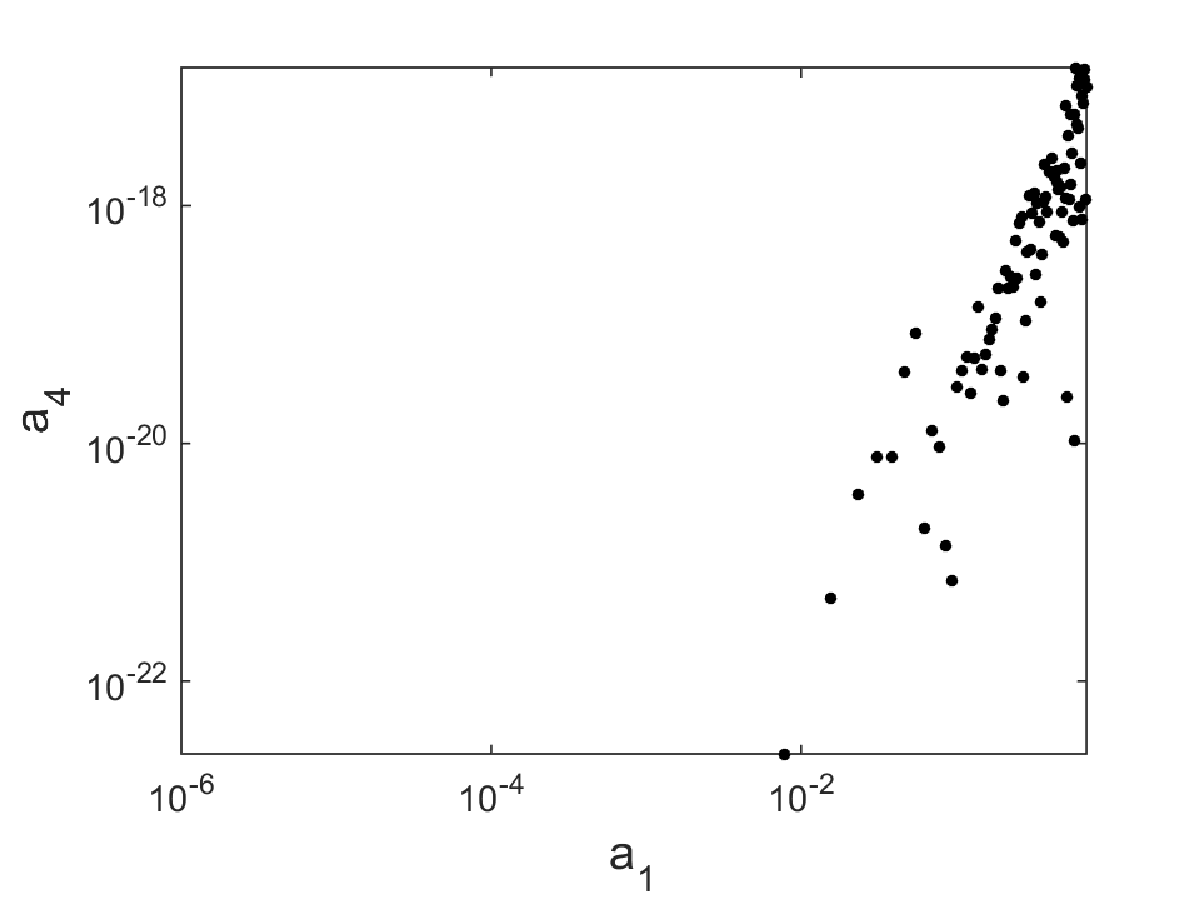}
\includegraphics[width=0.32\textwidth]{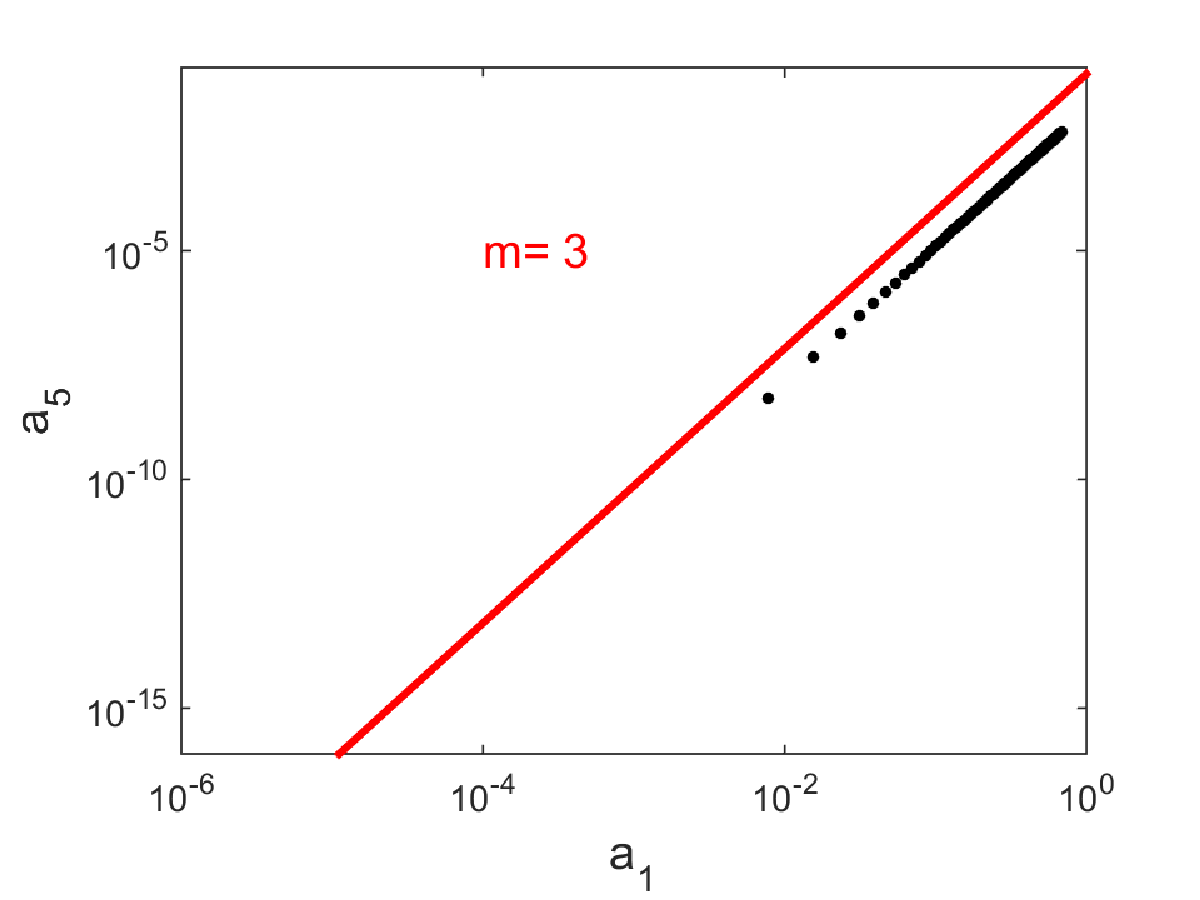}
\includegraphics[width=0.32\textwidth]{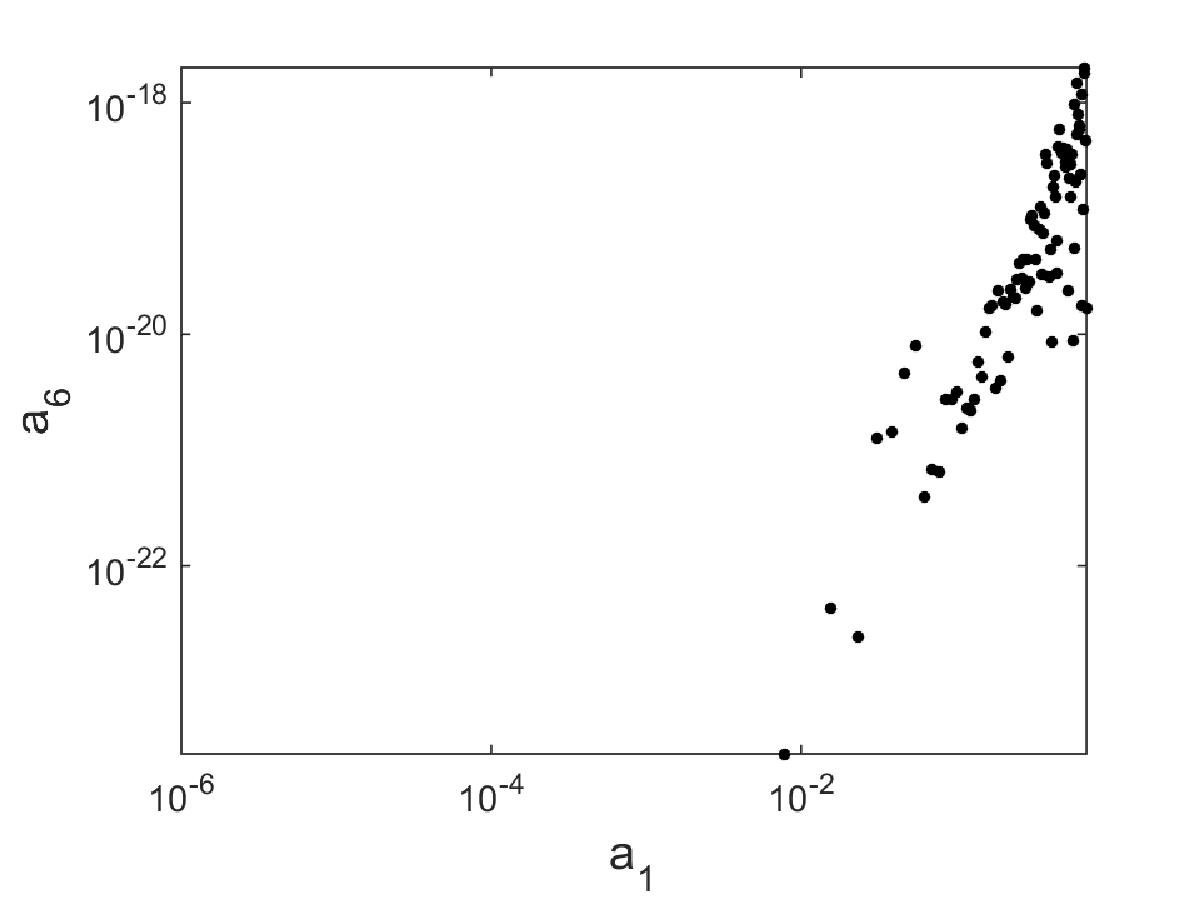}
\caption{Fourier coefficients of a branch of solutions of \eqref{eq:cubicSteadyState} with $p=2$ in the resonant regime with $K=3$, $\alpha = 1$ and $\beta = 1/10$. The top left figure is the bifurcation branch of $a_1$ as a function of $V$, followed by  log-log plots of $a_j$ vs. $a_1$, with a line of the labelled slope. We have that $a_1 = O(\epsilon)$ by assumption, and $a_3 = O(\epsilon)$. The solutions contain no even modes. \label{fig:mKdVModesbeta1_10}}
\end{figure}

We proceed as in Section \ref{sec:asymptotics}. We treat $a_1$ as a parameter and $(V, a_2, a_3, \hdots, a_N)^T$ as unknowns. We use a numerical continuation method by choosing a small value for $a_1$, computing a true solution using a Newton method. We scale the result and use it as an initial guess to compute a larger amplitude solution. This produces the bifurcation branches shown in the top-left corner of Figures \ref{fig:kawaModesbeta1_4} - \ref{fig:mKdVModesbeta1_10}. The wave profiles for the steady-state Kawahara equation \eqref{eq:cubicSteadyState} for the largest amplitude wave computed are shown in Figure \ref{fig:kawaharaProfiles} with $\beta = 1/4$ (left, non-resonant) and $\beta = 1/5$ (right, resonant).  The wave profiles for the steady-state modified fifth-order KdV equation \eqref{eq:cubicSteadyState} for the largest amplitude wave computed are shown in Figure \ref{fig:mKdVProfiles} with the non-resonant profile on the left with $\beta = 1/4$ and the resonant profile with $\beta = 1/10$ on the right. For the resonant profiles in these figures, we see that the main wave resembles a cosine, and there are smaller-amplitude oscillations of higher frequency, referred to as Wilton ripples in the context of water waves \cite{W15}.

To analyse how the coefficients $a_2, \hdots, a_N$ depend on $a_1$, we show log-log plots for resonant and non-resonant regimes with a linear fit and the slope of that fit labelled as $m$ on each plot of Figures \ref{fig:kawaModesbeta1_4} - \ref{fig:mKdVModesbeta1_10}. We consider the non-resonant case with $\beta = 1/4$ for the Kawahara equation ($p=1$) in Figure \ref{fig:kawaModesbeta1_4} and those of the modified fifth-order KdV equation ($p=2$) in Figure \ref{fig:mKdVModesbeta1_4}. The coefficients for the resonant $K=2$ mode with $\beta = 1/5$ are shown in Figure \ref{fig:kawaModesbeta1_5}. We see that $a_2$ grows like $O(\epsilon)$, in contrast to $O(\epsilon^2)$ in the non-resonant case. Note that for the resonant solutions, the bifurcation branch has a turning point. If we change which mode is resonant, e.g. $K=5$, by equating $\beta = 1/26$, we see that the ordering of the modes with respect to the powers of $\epsilon$ changes such that now $a_5=O(\epsilon^3)$ as shown in Figure \ref{fig:kawaModesbeta1_26}. This is similar to the result shown in \cite{HB88}. For the modified fifth-order KdV, only odd-index modes are resonant as shown in Figure \ref{fig:mKdVModesbeta1_10} where with $K=3$, $a_3$ is the dominant resonant coefficient.

\begin{figure}
\begin{center}
\includegraphics[width= 0.49\textwidth]{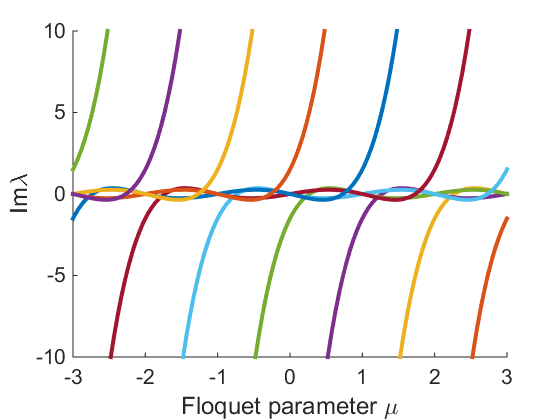}
\includegraphics[width= 0.49\textwidth]{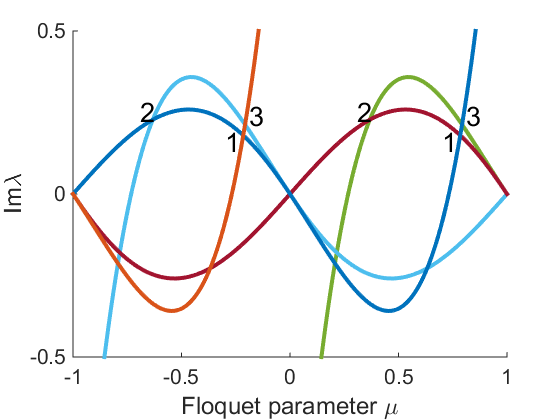}
\caption{The eigenvalues $\lambda_n^{\mu}$ given by \eqref{eq:smallAmpGenEval} for $\alpha = 1$ and $\beta = 1/4$. The panel on the right is a magnification near the horizontal axis of the panel on the left with the three unique collisions labelled.  \label{fig:evalsbeta1_4}}
\end{center}
\end{figure}

To analyse the stability of the obtained solutions, we compute the matrices \eqref{eq:firstOrderSeries} with $p=1$ for Kawahara and $p=2$ for modified fifth-order KdV. We check the number of Fourier modes needed to capture the instabilities by examining how many eigenvalue collisions exist for a given $\beta$. For example, with $\beta = 1/4$, there are 3 unique collisions as seen in Figure \ref{fig:evalsbeta1_4}. There we plot several eigenvalues $\lambda_n^{\mu}$, without restricting the Floquet parameter $\mu$ to be in $[-1/2, 1/2]$. Instead, we show the ``unfolded'' eigenvalues. Table \ref{tab:stabbeta1_4} gives the numerical values of the Floquet parameter $\mu$ for which these collisions occur. The signature column in the table refers to the Krein signature for the colliding eigenvalues.

\begin{figure}
\begin{center}
\includegraphics[width=0.49\textwidth]{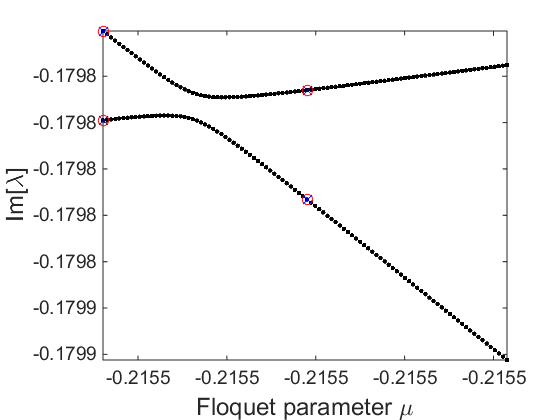}
\includegraphics[width=0.49\textwidth]{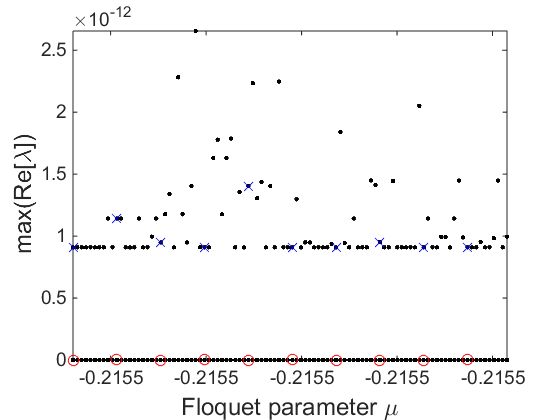}
\caption{The real and imaginary part of the spectrum  for $a_1=10^{-3}$ near the location of the collision labelled as 1 in Figure \ref{fig:evalsbeta1_4}. The theory predicts that this collision does not result in an instability, and this is verified here. On the left is Im($\lambda$) vs. the Floquet parameter $\mu$. On the right is Re($\lambda$). Red circles label every tenth point of the asymptotic prediction from the $2\times 2$ matrix given by \eqref{eq:beta1_4col1} and the blue crosses label the numerical computations for a larger $16 \times 16$ matrix. \label{fig:beta1_4col1}}
\end{center}
\end{figure}

\begin{figure}
\begin{center}
\includegraphics[width=0.49\textwidth]{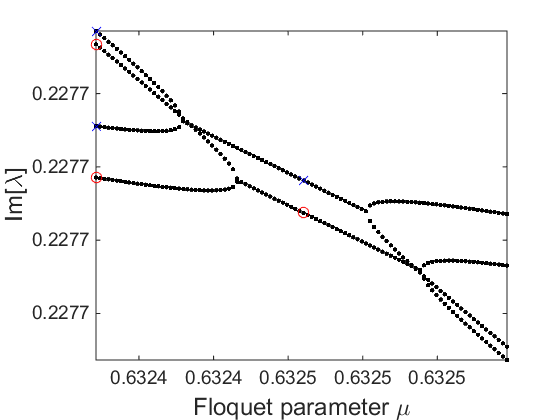}
\includegraphics[width=0.49\textwidth]{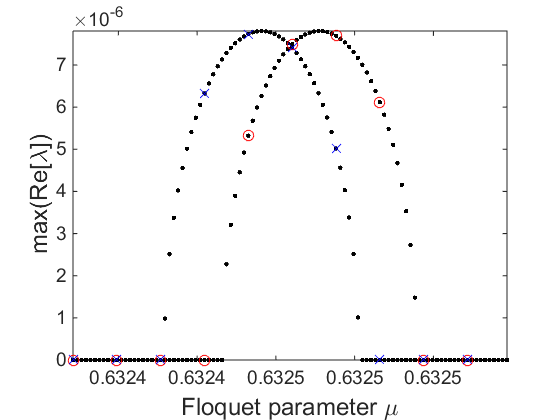}
\end{center}
\caption{The real and imaginary part of the spectrum  for $a_1=10^{-3}$ near the location of the collision labelled as 2 in Figure \ref{fig:evalsbeta1_4}. The theory predicts that this collision may result in an instability, and it is verified here that it does. On the left is Im($\lambda$) vs. the Floquet parameter $\mu$. On the right is Re($\lambda$). Red circles label every tenth point of the asymptotic prediction from the $2\times 2$ matrix given by \eqref{eq:beta1_4col2} and the blue crosses label the numerical computations for a larger $16 \times 16$ matrix. This illustrates that collisions resulting in higher-order growth rates require the consideration of matrices of larger size. 
\label{fig:beta1_4col2}}
\end{figure}

\begin{figure}
\begin{center}
\includegraphics[width=0.49\textwidth]{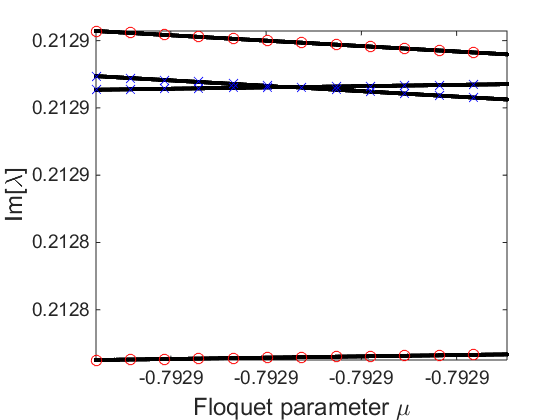}
\includegraphics[width=0.49\textwidth]{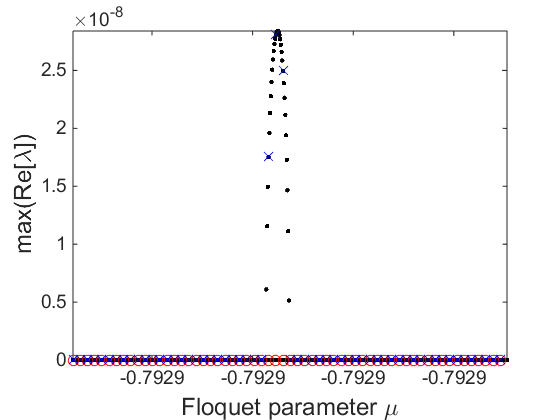}
\end{center}
\caption{
The real and imaginary part of the spectrum  for $a_1=10^{-3}$ near the location of the collision labelled as 3 in Figure \ref{fig:evalsbeta1_4}. The theory predicts that this collision may result in an instability, and it is verified here that it does. On the left is Im($\lambda$) vs. the Floquet parameter $\mu$. On the right is Re($\lambda$). Red circles label every tenth the asymptotic prediction from the $2\times 2$ matrix given by \eqref{eq:beta1_4col2} and the blue crosses label the numerical computations for a larger $16 \times 16$ matrix. This illustrates that collisions resulting in higher-order growth rates require the consideration of matrices of larger size.
\label{fig:beta1_4col3}}
\end{figure}

In order to see if instability arises as the amplitude increases, we compute the eigenvalues for the linearisation around a non-zero amplitude solution. We compare these numerically computed eigenvalues for a perturbation that contains many Fourier modes with the eigenvalues obtained analytically from the $2\times 2$ matrices for $\beta = 1/4$. The results are shown in Figures \ref{fig:beta1_4col1} - \ref{fig:beta1_4col3}. The analytically-obtained expressions of the eigenvalues are labelled with red circles and the numerical results are labelled with blue crosses. An avoided collision is shown in Figure~\ref{fig:beta1_4col1} for a small-amplitude solution with the numerical and asymptotic results in perfect agreement.  The numerical precision is O$(10^{-12})$, and the numerical results for the real part are showing zero, effectively. We see that for parameter regimes where the necessary condition for instability is met, this appears in the numerical results. As illustrated in Figures \ref{fig:beta1_4col2} and \ref{fig:beta1_4col3}, by comparing the eigenvalue collisions for solutions of nonzero amplitude using the perturbative calculation from the $2\times 2$ matrix and using the numerical results for a matrix containing many Fourier modes of the perturbation, we see that for $(m,n)$-collisions resulting in higher-order growth rates, it is necessary to consider matrices taking into account all modes between $m$ and $n$. Not doing so results in poor comparisons between the perturbative and numerical results for the value of the Floquet parameter for which the collision occurs.

\begin{figure}
\begin{center}
\includegraphics[width= 0.49\textwidth]{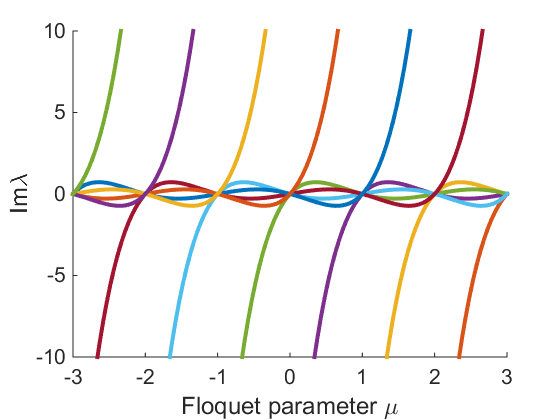}
\includegraphics[width= 0.49\textwidth]{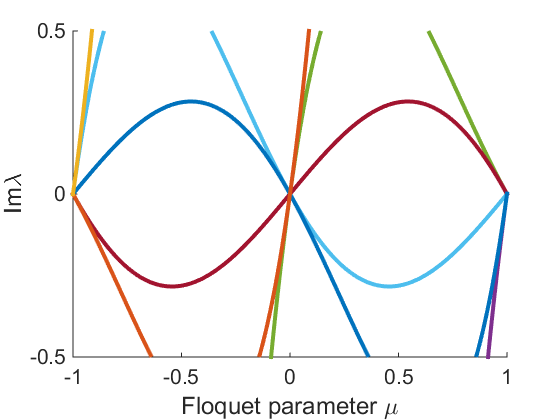}
\caption{The eigenvalues $\lambda_n^{\mu}$ given by \eqref{eq:smallAmpGenEval} for $\alpha = 1$ and $\beta = 1/5$ (resonant regime with $K=2$) with the plot on the right showing the region around the horizontal axis where the only collisions happens.\label{fig:evalsbeta1_5}}
\end{center}
\end{figure}

\begin{figure}
\begin{center}
\includegraphics[width = 0.34\textwidth]{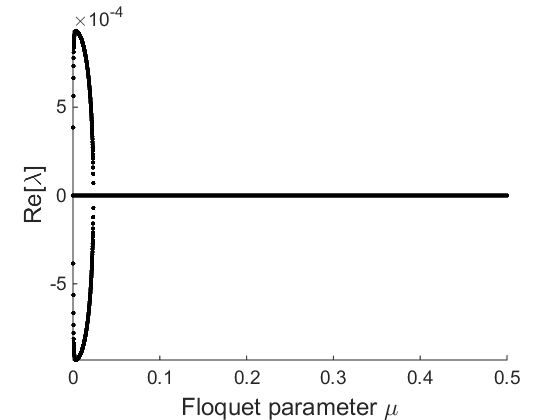}
\includegraphics[width = 0.34\textwidth]{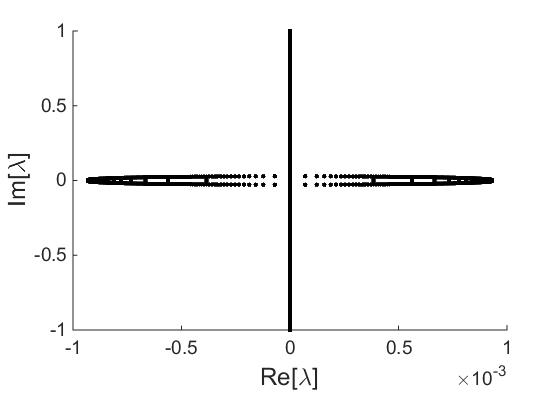}\\
\includegraphics[width = 0.34\textwidth]{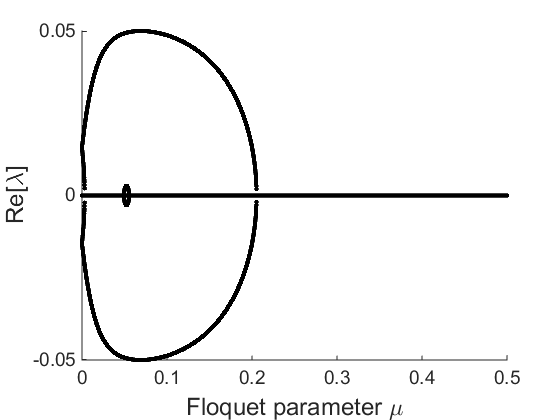}
\includegraphics[width = 0.34\textwidth]{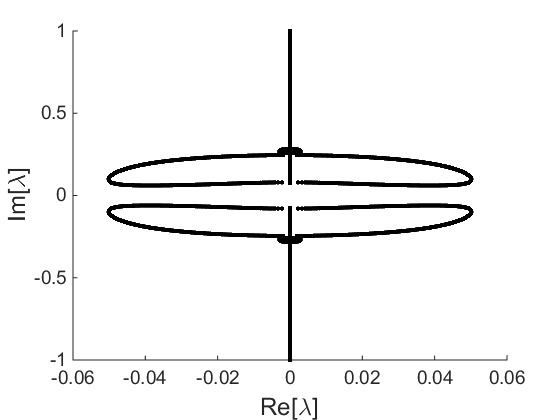}\\
\includegraphics[width = 0.34\textwidth]{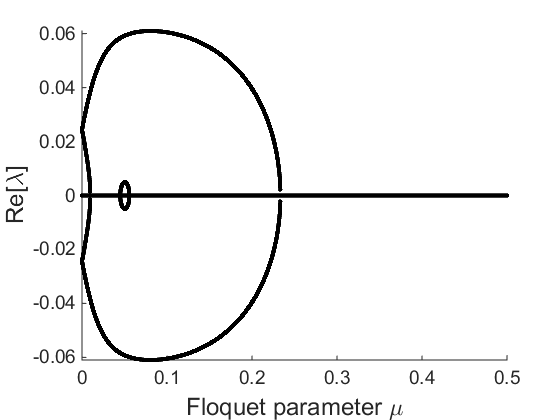}
\includegraphics[width = 0.34\textwidth]{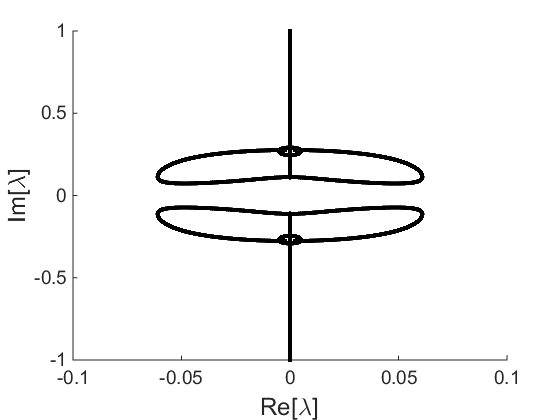}\\
\includegraphics[width = 0.34\textwidth]{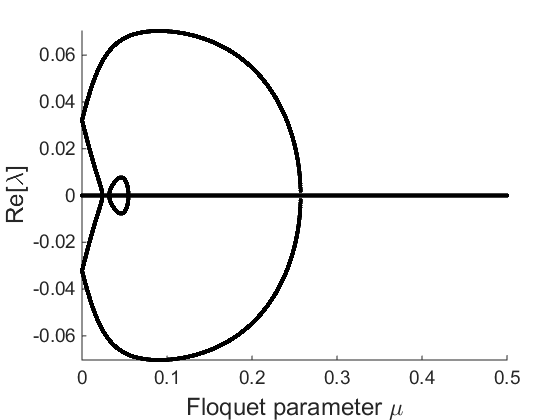}
\includegraphics[width = 0.34\textwidth]{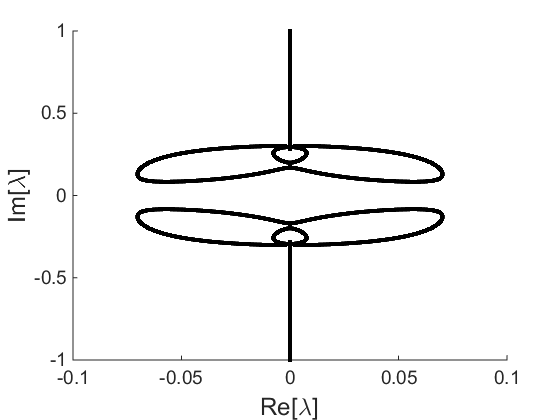}\\
\includegraphics[width = 0.34\textwidth]{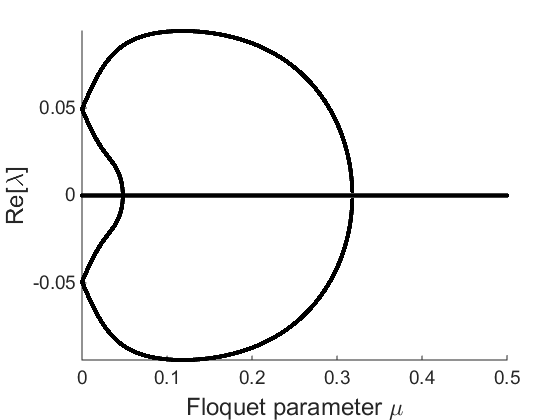}
\includegraphics[width = 0.34\textwidth]{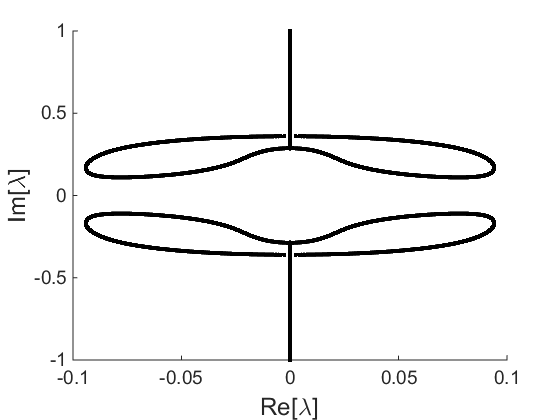}\\
\caption{Spectra for $\alpha = 1$ and $\beta = 1/5$ for which there is only one collisions at the origin for zero amplitude, as shown in Figure \ref{fig:evalsbeta1_5}. In the left column we show which perturbations (as determined by the Floquet parameter) are unstable with which growth rate. On the right the spectrum in the complex plane is displayed. Moving down, the amplitude of the solution increases. 
The observed instabilities originate from $\mu = 0$, move away from the origin and interact with each other.  \label{fig:spectrumbeta1_5}}
\end{center}
\end{figure}

Next we vary $\beta$, resulting in a different number of eigenvalue collisions. For example, in the resonant regime with the lowest resonant mode at $K=2$, many eigenvalues collide at the origin, seen in Figure \ref{fig:evalsbeta1_5}. However, for $K=2$ (and $\beta = 1/5$), there are collisions only for $\lambda=0$ and we cannot compute a definite signature. We examine how eigenvalues evolve as the amplitude is increased by examining the case with a single eigenvalue collision at the origin, as shown in Figure \ref{fig:spectrumbeta1_5}. We see (right column) that these eigenvalues move away from the origin and a second instability develops (more easily seen in the first column) where the resonant harmonic interacts with the perturbation.
These unstable eigenvalues interact and the graph of their spectrum becomes less ellipsoidal as amplitude increases. These results are easy to compute numerically, but the asymptotic expansions are unwieldy, emphasizing the benefit of the numerical approach.

\begin{table}
\begin{center}
\begin{tabular}{|c|c|c|c|c|}
\hline
$|m-n|$ & $\mu$ & Im($\lambda$) &Signature & Conclusion\\
\hline
1 & $0.7845$ & $-0.1798$ & same & stable \\
2 & $0.6324$ & $0.2277$ & different & instability possible \\
3 & $-0.7928$ & $0.2128$ & different & instability possible \\
\hline
\end{tabular}
\end{center}
\caption{The list of collisions obtained from solving \eqref{eq:quartic} with $\alpha=1$ and $\beta=1/4$. The collision label is equal to the order of the growth rate of the instability. The resulting numerical stability analysis is shown in Figures \ref{fig:beta1_4col1} - \ref{fig:beta1_4col3}. \label{tab:stabbeta1_4}}
\end{table}

\begin{figure}
\begin{center}
\includegraphics[width= 0.6\textwidth]{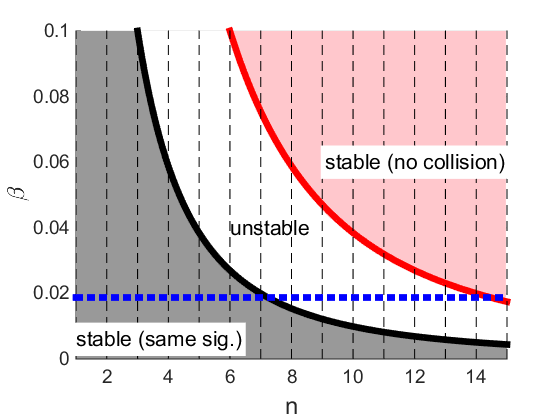}
\caption{Summary of the stability regions with the red line showing the value of $\beta = 3/160$ and the intersection at $|m-n|=14$. \label{fig:rootsbeta3_160}}
\end{center}
\end{figure}

Figure \ref{fig:rootsbeta3_160} is identical to Figure~\ref{fig:totInstab}, but using a value of $\beta = 3/160$ for the horizontal line determining the number of instabilities.  For this value there are 14 collisions, but only 7 with opposite Krein signature. In this case, we can summarise what happens for each eigenvalue collisions in Table \ref{tab:14Col}. This shows perfect agreement between our numerics and theory.

\begin{table}
\begin{center}
\begin{tabular}{|m{1.2cm}|m{1.2cm}|m{1.2cm}|m{2cm}|m{2cm}|m{2.2cm}|}
\hline
$|m-n|$ & $\mu$ & Im($\lambda$) &Signature & Conclusion & Numerics\\
\hline
1 & $5.09 $ & 62.82 & same & stable & \includegraphics[width= 0.13\textwidth]{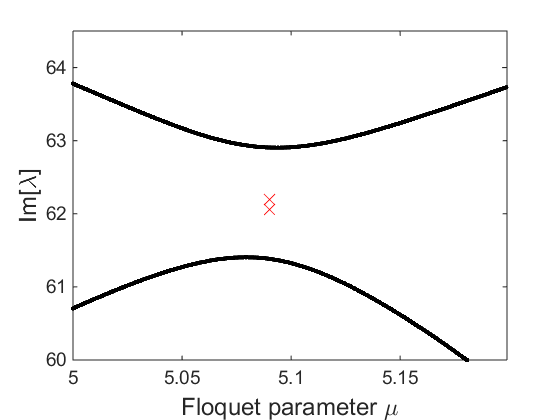} \\
2 & $5.48 $ & 51.62 &  same & stable & \includegraphics[width= 0.13\textwidth]{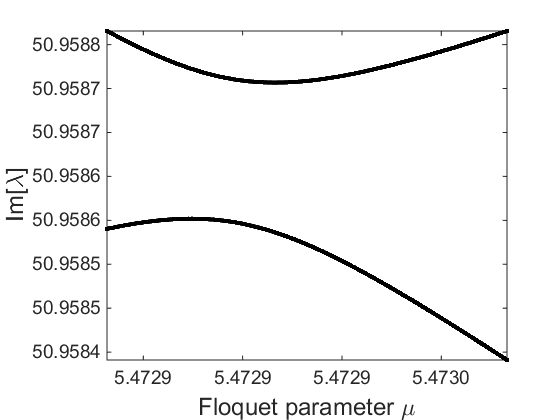} \\
3 & $4.79 $ & 35.98 & same & stable & \includegraphics[width= 0.13\textwidth]{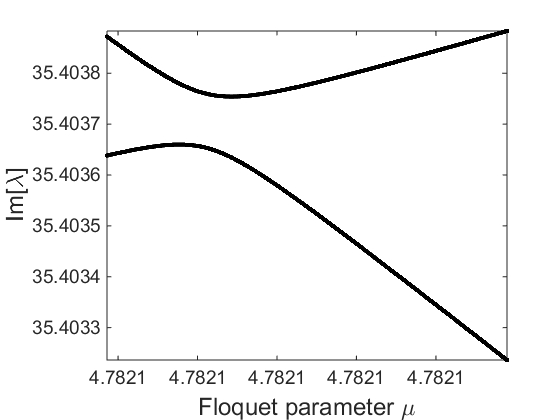}\\
4 & $5.02 $ & 19.78 & same & stable & \includegraphics[width= 0.13\textwidth]{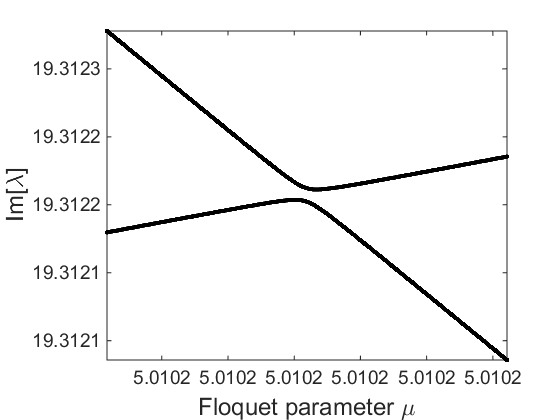}\\
5 & $4.16$ & 7.09 & same & stable & \includegraphics[width= 0.13\textwidth]{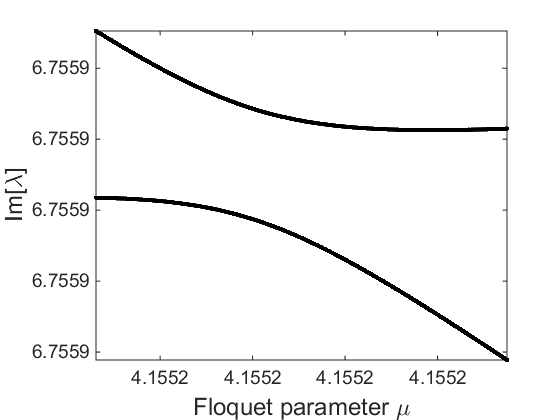}\\
6 & $4.23$ & 0.595 & same & stable& \includegraphics[width= 0.13\textwidth]{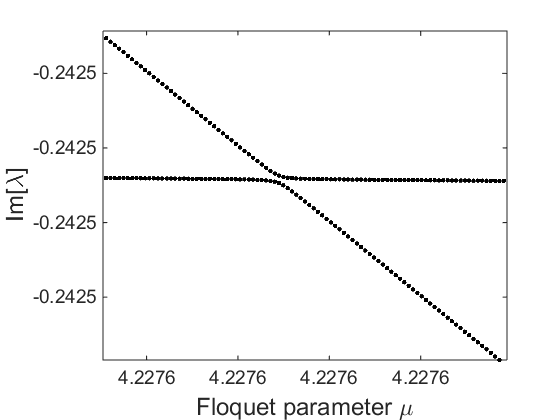}\\
7 & $3.24$ & -0.219 & same & stable & \includegraphics[width= 0.13\textwidth]{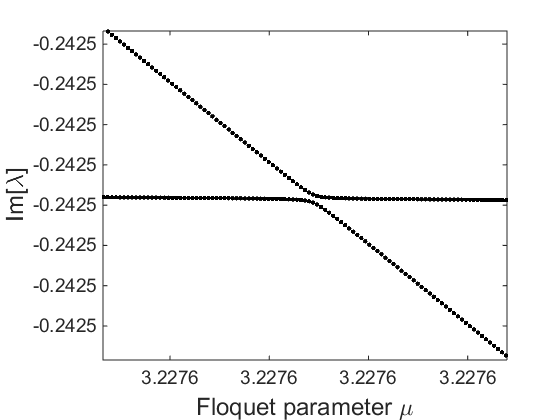}\\
8 & $-3.23$ & -0.305 & different & instability possible& \includegraphics[width= 0.13\textwidth]{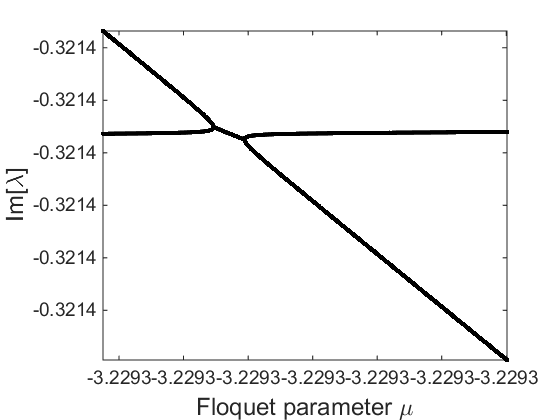}\\
9 & $2.27$ & -3.22 & different & instability possible& \includegraphics[width= 0.13\textwidth]{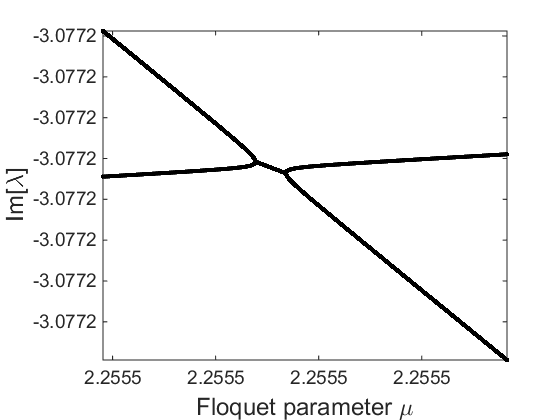}\\
10 & $2.36$ & -13.41 &  different & instability possible & \includegraphics[width= 0.13\textwidth]{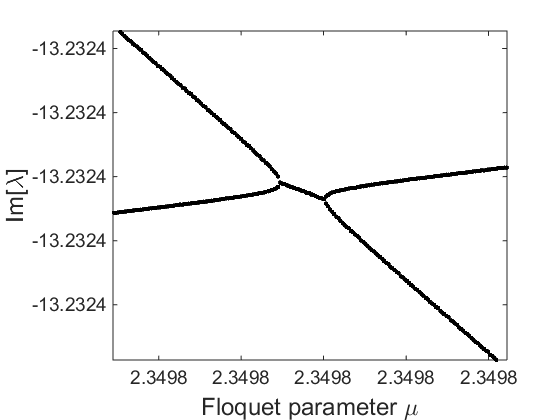}\\
11 & $1.49$ & -29.71& different & instability possible & \includegraphics[width= 0.13\textwidth]{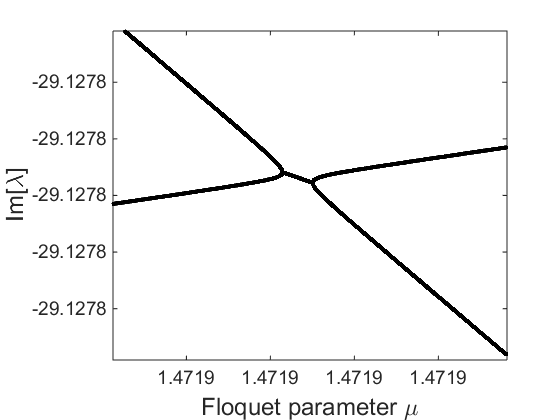}\\
12 & $1.64$ & -49.13 & different & instability possible & \includegraphics[width= 0.13\textwidth]{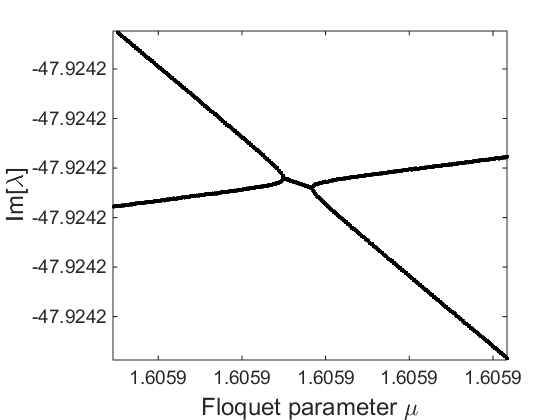}\\
13 & $0.74$ & -64.87 &different & instability possible & \includegraphics[width= 0.13\textwidth]{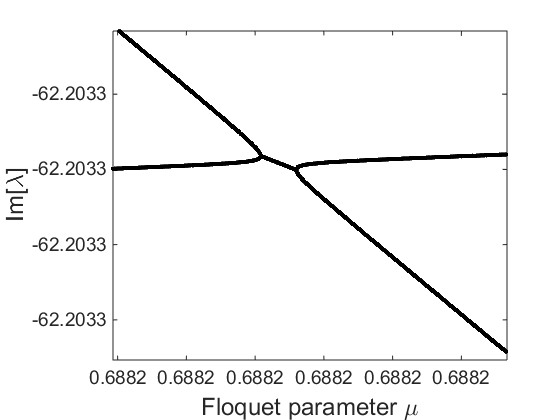}\\
14 & $0.69$ & -57.60 &different & instability possible & \includegraphics[width= 0.13\textwidth]{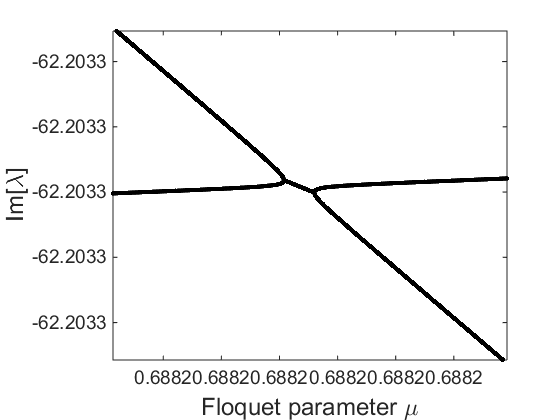}\\
\hline
\end{tabular}
\end{center}
\caption{The list of all collisions obtained from solving \eqref{eq:quartic} for $\alpha=1$ and $\beta=3/160$ with the numerical evaluations in the right column verifying the conclusions column. \label{tab:14Col}}
\end{table}

\section{Conclusion and Discussion}\label{sec:conclusion}

The main result of this work is the demonstration that small-amplitude stability criteria for the fifth-order KdV equation reduce to considering the roots of a quadratic equation. The full characterisation of the stability is given in Figure \ref{fig:totInstab}. By considering different forms of the nonlinearity, this analysis is extended to show how these instabilities depend on the small amplitude using a perturbation expansion of the solutions. These perturbation results are validated with numerical computations. The main criterion for having any possibility for an instability, is the existence of a generalised resonance as in \eqref{eq:resBeta}.

As is seen from the computations, even when the nonlinearity and the growth rate of the instability are small, we deviate from the asymptotic results as seen in Figures \ref{fig:beta1_4col2} and \ref{fig:beta1_4col3}. This implies it is necessary to proceed to higher order in the perturbation expansion and we may have to consider the perturbations of eigenvalues for the full matrix. Even at the order of our analysis, we see that the Floquet parameter for which an instability occurs changes even for small-amplitude solutions. This is why the instabilities are hard to find numerically, without any theoretical input. They are of small magnitude and appear for Floquet parameters different than the parameter for the collisions for zero-amplitude solutions. The perturbation expansion informs the numerical procedure and allows us to narrow down the parameter range where the instability can occur.

We may repeat this analysis for different dispersive equations by using the theory of \cite{KDT18}. The resulting polynomial equation determining the Krein signature will be of higher (than second) polynomial order. If only two dispersive terms are present, the bounds for where instabilities are found are given explicitly in \cite{KDT18}. If more terms are present, Sturm's theorem can be used to methodically determine how many negatively signed collisions exist depending on the parameters in the equation.

In this work, we have focussed on small solutions with small perturbations. This allowed us to keep very few terms in the perturbation expansion for the solutions, their perturbations and the growth rates. in general, more terms in the expansion can be kept. This leads to larger matrices. The perturbation of the eigenvalues of these matrices can be considered by using the theory available in \cite{Kato}. 
For fully nonlinear solutions, numerical computations remain the fastest and easiest way to obtain the full spectrum showing all the unstable perturbations.

\section{Acknowledgements}
We thank M. Haragus and D. Ambrose for insightful discussions.
This work was supported
by the Slovak Research and Development Agency under the contract No.~APVV-14-0378 (RK) and
generously supported by the National Science Foundation grant number NSF-DMS-1522677 (BD). Any opinions, findings, and conclusions or recommendations expressed in this material are those of the authors and do not necessarily reflect the views of the funding sources. We wish to thank Casa Mathem{\'a}tica Oaxaca,  the Erwin Schr\"{o}dinger Institute and Institute for Computational and Experimental Research in Mathematics (ICERM) for their hospitality during the development of the ideas for this work.

\bibliographystyle{plain}

\bibliography{mybib}

\begin{thebibliography}{10}

\bibitem{A15}
B.~F. Akers.
\newblock Modulational instabilities of periodic traveling waves in deep water.
\newblock {\em Physica D: Nonlinear Phenomena}, 300:26--33, 2015.

\bibitem{AG12}
B.~F. Akers and W.~Gao.
\newblock Wilton ripples in weakly nonlinear model equations.
\newblock {\em Communications in Mathematical Sciences}, 10(3):1015--1024,
  2012.

\bibitem{BDG02}
T.~J. Bridges, G.~Derks, and G.~Gottwald.
\newblock Stability and instability of solitary waves of the fifth-order {KdV}
  equation: a numerical framework.
\newblock {\em Physica D: Nonlinear Phenomena}, 172(1-4):190--216, 2002.

\bibitem{CJ17}
K.~M. Claassen and M.~A. Johnson.
\newblock Numerical bifurcation and spectral stability of wavetrains in
  bidirectional {W}hitham models.
\newblock {\em arXiv preprint arXiv:1710.09950}, 2017.

\bibitem{deconinck1}
B.~Deconinck and J.~N. Kutz.
\newblock Computing spectra of linear operators using the
  {F}loquet-{F}ourier-{H}ill method.
\newblock {\em Journal of Computational Physics}, 219:296--321, 2006.

\bibitem{DT15}
B.~Deconinck and O.~Trichtchenko.
\newblock High-frequency instabilities of small-amplitude solutions of
  {H}amiltonian {PDE}s.
\newblock {\em DCDS-A.~37}, pages 1323--1358, 2017.

\bibitem{HLS06}
M.~Haragus, E.~Lombardi, and A.~Scheel.
\newblock Spectral stability of wave trains in the {K}awahara equation.
\newblock {\em Journal of Mathematical Fluid Mechanics}, 8(4):482--509, 2006.

\bibitem{HB88}
S.~E. Haupt and J.~P. Boyd.
\newblock Modeling nonlinear resonance: a modification to the {S}tokes'
  perturbation expansion.
\newblock {\em Wave Motion}, 10(1):83--98, 1988.

\bibitem{HK08}
M.~Haˇraˇgu{\c{s}} and T.~Kapitula.
\newblock On the spectra of periodic waves for infinite-dimensional hamiltonian
  systems.
\newblock {\em Physica~D: Nonlinear Phenomena}, 237(20):2649--2671, 2008.

\bibitem{JZB10}
M.~A. Johnson, K.~Zumbrun, and J.~C. Bronski.
\newblock On the modulation equations and stability of periodic generalized
  {K}orteweg--de {V}ries waves via {B}loch decompositions.
\newblock {\em Physica D: Nonlinear Phenomena}, 239(23-24):2057--2065, 2010.

\bibitem{kapituladeconinck}
T.~Kapitula and B.~Deconinck.
\newblock On the spectral and orbital stability of spatially periodic
  stationary solutions of generalized {K}orteweg-de {V}ries equations.
\newblock In {\em Hamiltonian partial differential equations and applications},
  pages 285--322. Springer, 2015.

\bibitem{Kato}
T.~Kato.
\newblock {\em Perturbation theory for linear operators}.
\newblock Classics in Mathematics. Springer-Verlag, Berlin, 1995.

\bibitem{K72}
T.~Kawahara.
\newblock Oscillatory solitary waves in dispersive media.
\newblock {\em J. Phys. Soc. Jpn.}, 33:1015--1024, 1972.

\bibitem{krein2}
M.~G. Kre{\u\i}n.
\newblock On the application of an algebraic proposition in the theory of
  matrices of monodromy.
\newblock {\em Uspehi Matem. Nauk (N.S.)}, 6(1(41)):171--177, 1951.

\bibitem{MS86}
R.~S. MacKay and P.~G. Saffman.
\newblock Stability of water waves.
\newblock In {\em Proc. R. Soc. Lond. A}, volume 406, pages 115--125. The Royal
  Society, 1986.

\bibitem{M82}
J.~W. McLean.
\newblock Instabilities of finite-amplitude water waves.
\newblock {\em Journal of Fluid Mechanics}, 114:315--330, 1982.

\bibitem{MGK68}
R.~M. Miura, C.~S. Gardner, and M.~D. Kruskal.
\newblock Korteweg-de vries equation and generalizations. ii. existence of
  conservation laws and constants of motion.
\newblock {\em Journal of Mathematical physics}, 9(8):1204--1209, 1968.

\bibitem{KDT18}
B.~Deconinck R.~Koll{\'a}r and O.~Trichtchenko.
\newblock Direct characterization of spectral stability of small amplitude
  periodic waves in scalar {H}amiltonian problems via dispersion relation.
\newblock {\em submitted for publication}.

\bibitem{SW02}
G.~Schneider and C.~E. Wayne.
\newblock The rigorous approximation of long-wavelength capillary-gravity
  waves.
\newblock {\em Archive for rational mechanics and analysis}, 162(3):247--285,
  2002.

\bibitem{TDW16}
O.~Trichtchenko, B.~Deconinck, and J.~Wilkening.
\newblock The instability of {W}ilton ripples.
\newblock {\em Wave Motion}, 66:147--155, 2016.

\bibitem{W15}
J.R. Wilton.
\newblock On ripples.
\newblock {\em Philosophical Magazine Series 6}, 29(173):688--700, 1915.

\end{thebibliography}

\end{document}